\documentclass[acmtog, screen, nonacm]{acmart}

\usepackage{algorithm}
\usepackage{algpseudocode}
\usepackage{enumitem}
\usepackage{placeins}

\algblockdefx[ParallelFor]{ParallelFor}{EndFor}[1]{\textbf{parallel for} #1 \textbf{do}}{\textbf{end for}}

\AtBeginDocument{%
  }

\begin{document}

\title{Blue Noise as a Lattice Gibbs Ensemble}

\author{Zhuoran Yi}
\affiliation{%
  \institution{University of Utah}
  \city{Salt Lake City}
  \country{USA}}
\email{zhuoranyi@siggraph.org}

\begin{abstract}
We present an algorithm for generating blue-noise point sets that is simultaneously local and parallel, and whose output is a draw, up to a bounded error, from an explicit probability distribution. Methods that optimize point positions achieve high spectral quality but tend to couple all points globally, while tile-based methods, the main existing local approach, assemble their output from point sets constructed offline rather than generating it from scratch; we show that locality does not require giving that up. Concretely, we model blue noise as a Gibbs distribution over a lattice, with a pairwise penalty whose continuous parameters control repulsion strength, scale, and hardness. To sample this distribution locally, we trace each site's dependencies backward in time and truncate the trace to a bounded region known before sampling begins; tiles can therefore be generated independently in any order, with no communication between them. We stipple a 14557$\times$8418 James Webb Space Telescope image tile by tile, with peak memory under 51 MiB, set by the working window rather than the image dimensions. Spectral quality is competitive with existing methods at matched density, and the construction extends naturally to multi-class outputs.
\end{abstract}



\begin{teaserfigure}
    \centering
    \includegraphics{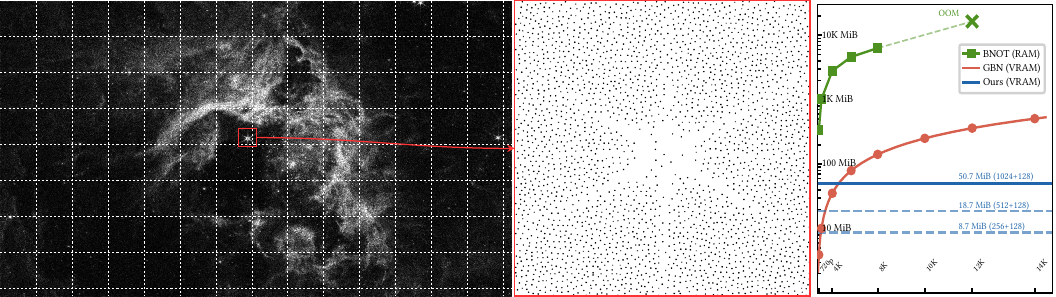}
    \vspace{-2em}
    \caption{\textbf{Adaptive stippling of the Tarantula Nebula}~\cite{WebbTarantula2022}, produced by our blue-noise sampler. Stippling reproduces an image using only dots, varying their local density to match brightness. The dots must also stay irregular and evenly spaced, so that no clumps or grid patterns appear. This is the blue-noise property. \emph{Left}: the full result at $1{,}201{,}764$ points. White dashed lines mark the boundaries of the $1024\times1024$ core tiles the image was generated in; the output is identical to generating the whole image at once. The dot radius is chosen so that the dots would just fill the image if spread uniformly. Local coverage then tracks local density, so the stipple reproduces the gray levels of the original. \emph{Center}: a crop of the red box, where the sampler tracks brightness from dense ionized filaments to sparse dark dust lanes. \emph{Right}: peak memory versus image resolution. Gaussian blue noise (GBN) and Blue Noise through Optimal Transport (BNOT), two state-of-the-art blue-noise methods, both build the point set globally, so their memory grows with the image and BNOT fails at $12$K with a 16 GiB memory limit. Our sampler works locally, so its memory is set by the working window (tile plus halo) alone and does not grow with the image.}
    \label{fig:teaser}
\end{teaserfigure}

\maketitle

\section{Introduction}
Blue noise refers to point sets that are irregular but evenly spaced. Regular grids alias and fully random points clump, so blue noise sits between the two \cite{Dippe1985Antialiasing,Cook1986StochasticSI}. Rendering, stippling, remeshing, and simulation all rely on it for this reason. The two requirements pull against each other, which is what makes blue noise hard to generate.

Methods for generating blue-noise point sets broadly follow either a \emph{Lagrangian} or an \emph{Eulerian} viewpoint. Lagrangian methods represent a point set\footnote{In this paper a \emph{point} is a single dot (an occupied site in the lattice model) and a \emph{pattern} is one complete output. We use ``sample'' only as a verb, and in ``sampler'' and ``sampling''.} as particle positions in continuous space. Optimization-based methods move those positions toward a target configuration. Sequential stochastic methods, such as dart throwing and its many Poisson-disk variants, place points one at a time and reject candidates that violate a minimum-distance constraint. Eulerian methods represent a point set as a binary occupancy field on a discrete lattice. Void-and-cluster methods swap occupied and unoccupied sites to improve uniformity. Tile-based methods are complementary to all of these: they are fast and support spatially varying density~\cite{Kopf2006RecursiveWangTiles,Wachtel2014FastTileBased}, but their output is assembled from point sets constructed offline rather than sampled.

The two families leave different gaps. Many Lagrangian methods couple point positions through a global density estimate, a global optimization objective, or a global transport plan. Sequential Lagrangian methods instead make each placement depend on earlier accepted points. Both cases make a provable finite dependency radius difficult to obtain. They also make tiled or streaming execution difficult to certify. Eulerian representations support local and parallel execution. However, the Eulerian methods described above define procedures rather than explicit probability distributions. Their statistical properties are therefore accessible only empirically, and repulsion strength is not exposed as a continuous parameter of a well-defined ensemble.

To close both gaps, we take the lattice as the model itself and define a Gibbs distribution, in which a pattern's probability falls exponentially with an energy that sums over pairs of nearby occupied sites. Both ingredients exist in prior work but not together: \citet{Wei2008ParallelPD} used a grid to schedule a Lagrangian method, and \citet{Fattal2011BluenoisePS} used a Gibbs distribution over continuous point coordinates. The pairwise form means distant interactions can be dropped with a change to the distribution we can bound. Either alone is not enough: a global energy on a lattice still couples everything, and a pairwise energy on continuous coordinates has no fixed neighborhood to exploit.

Sampling this ensemble normally means running a Markov chain forward until it has mixed (forgotten its starting state), with no bound on cost. Instead we trace the chain backward, following Coupling Towards The Past (CTTP) \cite{Feng2023Derandomising}, and cut the trace at a fixed depth. Truncation costs accuracy, bounded and decaying with depth, but fixes each site's dependency region in advance, so a tile reproduces exactly what the same sampler would produce there on any larger domain, without communication between tiles. Figure~\ref{fig:teaser} shows a 14K stipple generated tile by tile in under 51 MiB, bit-identical to a single pass over the whole field. What the figure demonstrates is streaming: tiles can be produced in any order, and memory does not grow with the input.

In summary, our main contributions are:
\begin{enumerate}[]
    \item We generate blue noise by sampling a lattice Gibbs model with a pairwise potential, and show empirically that the realizations are competitive in quality with existing methods, including optimization-based ones, while their spectral character responds to the model's continuous parameters.

    \item Starting from the backward viewpoint of CTTP~\cite{Feng2023Derandomising,liu2026local}, we present, to our knowledge, the first sampler for an explicit blue-noise distribution that is simultaneously local and parallel under arbitrary boundary conditions on the plane, together with a certified bound guaranteeing that tiled and full-domain generation agree exactly.

    \item We present a tiled adaptive-sampling algorithm for output sizes far beyond what a single pattern could otherwise reach, whose tiled output is bitwise identical to a single-pass full-domain reference.
\end{enumerate}

The remainder of the paper is organized as follows. Section~\ref{sec:related} reviews prior blue-noise methods and the statistical-mechanics and exact-sampling literature we build on. Section~\ref{sec:model} defines the ensemble and its parameters. Section~\ref{sec:sampler} derives the backward sampler and its truncations, stating the error and locality guarantee that each one carries. Section~\ref{sec:experiments} evaluates the model and sampler and demonstrates adaptive and multi-class extensions. Section~\ref{sec:conclusion} discusses scope and future work. Proofs are in the supplementary material.

\section{Background and Related Work}
\label{sec:related}

\subsection{Blue-Noise Generation in Graphics}

\paragraph{Moving points around.}
One family starts from a random point set and moves the points until they are well spaced. Lloyd relaxation~\cite{Lloyd1982LeastSQ} does this by repeatedly recentering each point in its Voronoi cell, and weighted Voronoi stippling~\cite{Secord2002WeightedVS} is the classic graphics application. Later work replaced the objective. Capacity-constrained Voronoi and Delaunay constructions equalize the area or mass assigned to each point \cite{Balzer2009CapacityconstrainedPD,Xu2011CapacityConstrainedDT}. Farthest-point optimization directly increases the minimum distance between points~\cite{Schlmer2011FarthestpointOP}, while variational blue-noise sampling relaxes capacity-constrained Voronoi partitions into a continuous energy~\cite{Chen2012VariationalBN}. Blue Noise through Optimal Transport (BNOT) formulates blue-noise generation as a capacity-constrained transport problem over power diagrams \cite{Goes2012BlueNT}; centroidal power diagrams provide efficient solvers for related optimal-partition problems \cite{Xin2016CentroidalPD}; and sliced optimal transport (SOT) brings optimal-transport point-set generation to higher dimensions \cite{Paulin2020SlicedOT}. Gaussian blue noise (GBN) directly minimizes a Gaussian-kernel energy and extends smoothly to high-dimensional and adaptive generation~\cite{Ahmed2022GaussianBN}. Electrostatic halftoning simulates repulsive electrostatic forces \cite{Schmaltz2010ElectrostaticHalftoning,Gwosdek2011FastEH}, while particle-based relaxation uses SPH or N-body dynamics to obtain related blue-noise configurations \cite{Jiang2015BlueNS,Wong2017BlueNS}.

These methods give the best spectra available today. They also share a structure: each point's final position depends on where all the others end up. The optimization cannot be split into independent pieces, so the whole point set has to be held and updated together, and what comes out is one good configuration rather than a pattern drawn from a distribution.

\paragraph{Placing points one at a time.}
A second family places points sequentially and rejects any candidate that lands too close to an existing one. Dart throwing \cite{Dippe1985Antialiasing,Cook1986StochasticSI} is the original form: propose a uniformly random location, accept it if no accepted point is within the conflict radius, repeat. Most later work makes this faster, or maximal, rather than changing the rule. Spatial data structures and scalloped-sector representations accelerate Poisson-disk generation \cite{Dunbar2006ASD}, while Voronoi-based algorithms and grid-based linear-time methods provide alternative efficient constructions \cite{Jones2006PoissonDisk,Bridson2007FastPD}. Parallel algorithms and accurate multidimensional dart throwing extend the same construction across processors and dimensions \cite{Wei2008ParallelPD,Gamito2009AccurateMD}. Efficient maximal Poisson-disk algorithms, gap-processing methods, and spoke-darts constructions address maximality, adaptivity, and high-dimensional generation \cite{Ebeida2011MaximalPD,Yan2013GapProcessing,Mitchell2018SpokeDarts}. A different route starts from far too many candidates and prunes them down~\cite{Yuksel2015SampleElimination}.

The accept-or-reject rule is what makes these methods fast, and also what limits them. A candidate is either in or out. Parameters can move the conflict radius, but not how strongly points repel at a given distance, and the resulting correlation structure is whatever the rule happens to produce. Placement also depends on which points were accepted earlier, which ties each decision to a history rather than to a bounded neighborhood.

\paragraph{Working on a lattice.}
A third family drops continuous coordinates and works with a binary occupancy field on a grid. Void-and-cluster \cite{Ulichney1993VoidandclusterMF} swaps occupied and empty sites until the pattern looks uniform, and is still widely used for dither masks. It runs on a lattice but not locally: each swap needs the
largest void and the tightest cluster in the whole field, so every step looks at everything.

\paragraph{Distributing precomputed point sets.}
Tile-based methods, in contrast, run locally~\cite{Kopf2006RecursiveWangTiles}. They generate a library of point sets offline and stitch arbitrarily large outputs from it at run time. Wang-tile methods populate a finite set of square tiles with precomputed point sets and tile the plane by matching edge colors~\cite{Cohen2003WangTiles}. Recursive variants use a self-similar substitution structure, allowing arbitrarily large areas and arbitrarily fine local detail to be unrolled from the same small set of base tiles and substitution rules \cite{Ostromoukhov2004FastHierarchical,Kopf2006RecursiveWangTiles}. Later work extends self-similar tiling to polyomino shapes and, with polyhex tiles, to user-specified target spectra \cite{Ostromoukhov2007Polyominoes,Wachtel2014FastTileBased}. This is the main existing route to output whose size is not bounded by memory, so it is worth being precise about what happens at run time. The arrangement of tiles is determined by matching or substitution rules over a finite set of stored point sets. No point is generated from scratch at run time. Which local arrangements can appear is fixed offline. Blue-noise output of arbitrary size can thus be reached in two broad ways: construct point sets offline and distribute them, as tile-based methods do, or generate the output directly at run time. The two are complementary, since the point sets must be produced by some offline procedure~\cite{Wachtel2014FastTileBased}. Our sampler can in principle serve as that procedure, because it supports arbitrary boundary conditions, and it needs no precomputed library: a $256\times256$ pattern takes $13.0$~ms at our low-budget setting and $155.6$~ms at our high-budget setting on a laptop GPU (Section~\ref{sec:experiments}).

\subsection{Explicit Distributions Over Point Sets}

All the methods above produce point sets, but none of them writes down the distribution those point sets are drawn from. A notable exception in graphics is Fattal's blue-noise model~\cite{Fattal2011BluenoisePS}, which defines an explicit Boltzmann--Gibbs distribution over point coordinates and samples it with a multiscale Langevin--Metropolis scheme. The target is a distribution, not a single good configuration, and randomness is part of the object rather than a construction device. Pair-correlation synthesis~\cite{Oztireli2012AnalysisAS} comes from the statistics side: it generates point sets whose pair correlation function matches a target, possibly taken from an example set, using a generalized dart throwing followed by gradient-descent fitting. What it specifies is a statistic of the output, not the probability law from which the output is drawn.

We take the same view and differ in how the energy is built. Fattal's energy compares a kernel density estimate of the pattern against a target density, so each point's contribution depends on the field that the surrounding points create, and with the $L_1$ norm used in practice it is not a sum of pair terms. Ours is a sum of independent pair terms. The two choices suit different goals: a density-matching energy is natural for prescribing a spatially varying density, while a pairwise energy on a lattice gives closed-form single-site conditionals (Section~\ref{sec:model-local}), which the backward sampler of Section~\ref{sec:sampler} needs; the conditionals of Fattal's model are not available in closed form~\cite{Fattal2011BluenoisePS}.

The idea of describing repulsive point patterns by a Gibbs density is older than its use in graphics. The Strauss process \cite{Strauss1975AMF} is the continuous-space process whose pair potential is the $p\to\infty$ limit of the one we use, a step function, and \citet{MollerWaagepetersen2004} treat this literature, including exact simulation by dominated CFTP. What differs here is the discretization: on a lattice the process becomes a Markov random field of bounded degree, which is what the sampler in Section~\ref{sec:sampler} needs.

\subsection{Backward Sampling}

Sampling from a Markov chain's stationary distribution normally means running it forward and stopping once it has mixed, which requires guessing how long that takes. Coupling From the Past \cite{Propp1996Exact} avoids the guess: it runs coupled chains from every possible starting state, backward from a remote past, and stops when they agree. Coupling Towards The Past \cite{Feng2023Derandomising,liu2026local} narrows this to a single variable, revealing only the randomness that variable's own history depends on rather than simulating the whole chain.

Our sampler follows the same backward idea but cuts the history at a fixed depth. Exactness is given up in exchange for a cost and a dependency region that are both known before sampling starts, which is what tiled execution needs.

\section{A Lattice Model of Blue Noise}
\label{sec:model}

Blue noise means points that are irregular but evenly spaced. In the graphics sense, blue noise means stochastic patterns with controlled density, reduced short-range clustering, suppressed low-frequency power after removing the mean, and low directional artifacts beyond the underlying discretization. These are properties of a pattern's statistics, not of any individual pattern, which is why we define blue noise here as a probability distribution over point patterns rather than as the output of a procedure.

The model produces such a distribution by penalizing pairs of nearby occupied sites. The penalty is finite, so close pairs are discouraged but not forbidden and the pattern retains stochastic variation. Its strength, spatial scale, and hardness are parameters of the distribution, so changing them moves the output statistics continuously rather than switching to a different construction. The resulting spectrum and pair-correlation curve emerge as statistics of the drawn patterns rather than being prescribed.

Section~\ref{sec:sampler} gives the sampler for this model.

\subsection{The Model}
\label{sec:model-lattice}

Let $\Lambda\subset\mathbb{Z}^2$ be a finite lattice domain. Each site $i\in\Lambda$ carries a binary occupancy variable $v_i\in\{0,1\}$, with $v_i=1$ meaning a point sits at that site. A configuration is $v=(v_i)_{i\in\Lambda}\in\Omega_\Lambda$ with $\Omega_\Lambda=\{0,1\}^{\Lambda}$, and we write $N_\Lambda(v)=\sum_{i\in\Lambda}v_i$ for the point count and $\rho_\Lambda(v)=N_\Lambda(v)/|\Lambda|$ for the empirical density.

The kernel below has infinite support, so sites near the edge of $\Lambda$ interact with sites outside it and a finite domain needs a boundary condition. Write $\tau\in\{0,1\}^{\Lambda^c}$ for an assignment of occupancies on $\Lambda^c$, the sites outside the domain, and
\begin{equation}
    v_i^\tau =
    \begin{cases}
        v_i, & i\in\Lambda,\\
        \tau_i, & i\notin\Lambda,
    \end{cases}
    \label{eq:extended-config}
\end{equation}
for the resulting extended configuration. The model is defined for any $\tau$. Which convention to use, such as an empty exterior, periodic wrapping, or a randomized one, is a choice made by the sampler rather than part of the model; Section~\ref{sec:sampler-dp} returns to it.

Two occupied sites repel each other by an amount that depends only on how far apart they are. We use a generalized Gaussian profile on lattice offsets $d\in\mathbb{Z}^2\setminus\{0\}$,
\begin{equation}
    k_{\sigma,p}(d)
    =
    \exp\!\left[-\left(\frac{\|d\|_2}{\sigma}\right)^p\right],
    \label{eq:kernel}
\end{equation}
where $\sigma>0$ sets the distance over which sites interact and $p>0$ sets how sharply the repulsion falls off. This assigns a symmetric weight $K_{ij}=k_{\sigma,p}(j-i)=K_{ji}>0$ to every pair of distinct sites, and the repulsive energy of a configuration is the sum over all pairs with at least one site inside $\Lambda$,
\begin{equation}
    P_\Lambda^\tau(v)
    =
    \sum_{\substack{\{i,j\}\subset\mathbb{Z}^2\\
                    i\ne j,\ \{i,j\}\cap\Lambda\neq\emptyset}}
    K_{ij}\, v_i^\tau v_j^\tau .
    \label{eq:pair-energy}
\end{equation}
The kernel has finite lattice tail mass for every $p>0$, so this sum converges. Pairs lying entirely outside $\Lambda$ are omitted because they take the same value for every $v$ and cancel in the normalization below.

Adding a term that rewards occupancy gives the energy
\begin{equation}
    H_{\alpha,\beta}^{\Lambda,\tau}(v)
    =
    -\alpha N_\Lambda(v)
    +
    \beta P_\Lambda^\tau(v),
    \label{eq:hamiltonian}
\end{equation}
where $\alpha\in\mathbb{R}$ controls how strongly occupancy is favored and $\beta\ge 0$ how strongly close pairs are penalized.\footnote{A conventional thermodynamic parameterization would use a chemical potential $\mu$, a pair strength $J$, and a temperature $T$. The distribution depends only on $\alpha=\mu/T$ and $\beta=J/T$, so $J$ and $T$ are not separately identifiable from the probability law and we work with the ratios directly.} Because $\beta P_\Lambda^\tau$ is finite for every configuration, close points are discouraged but never forbidden.

The distribution over configurations is then
\begin{equation}
    \mu_{\Lambda}^{\tau}(v)
    = \frac{1}{Z_{\Lambda}^{\tau}}
      \exp\!\left[
        \alpha\sum_{i\in\Lambda}v_i
        - \beta\!\!\!\sum_{\substack{\{i,j\}\subset\mathbb{Z}^2\\
                        i\ne j,\ \{i,j\}\cap\Lambda\neq\emptyset}}\!\!\!
          K_{ij}\, v_i^\tau v_j^\tau
      \right],
    \label{eq:gibbs}
\end{equation}
where $Z_{\Lambda}^{\tau}$ sums the exponential over all configurations in $\Omega_\Lambda$. Each close pair of occupied sites multiplies a configuration's probability by $e^{-\beta K_{ij}}$, a factor below one that shrinks as the pair gets closer, so raising $\beta$ makes clumps progressively less likely. At $\beta=0$ the penalties vanish and the sites become independent Bernoulli variables, each occupied with probability $1/(1+e^{-\alpha})$. We write $\mu_{\Lambda}^{\tau}(v)\propto\exp[-H_{\alpha,\beta}^{\Lambda,\tau}(v)]$ for the compact form and use it from here on.

\subsection{Local Conditional Probabilities}
\label{sec:model-local}

The normalizing constant $Z_{\Lambda}^{\tau}$ sums over all $2^{|\Lambda|}$ configurations and is out of reach for any domain of interest. The conditional law of a single site, however, is available in closed form, and it is all the sampler of Section~\ref{sec:sampler} ever needs.

For a site $i\in\Lambda$, define its local repulsive field
\begin{equation}
    S_i(v^\tau)
    =
    \sum_{j \in \mathbb{Z}^2\setminus\{i\}}
    K_{ij}\, v_j^\tau ,
    \label{eq:local-field}
\end{equation}
the weighted count of occupied sites seen from $i$. Turning $v_i$ from $0$ to $1$ leaves every term of $H_{\alpha,\beta}^{\Lambda,\tau}$ unchanged except the ones involving $i$, so the energy changes by $\Delta H_i = -\alpha + \beta S_i(v^\tau)$, and the conditional probability of occupying the site is
\begin{equation}
    \mu_{\Lambda}^{\tau}
    \left(
        v_i=1
        \mid
        v_{\mathbb{Z}^2\setminus\{i\}}
    \right)
    =
    \operatorname{Sigmoid}
    \left(
        \alpha-\beta S_i(v^\tau)
    \right),
    \label{eq:local-conditional}
\end{equation}
with $\operatorname{Sigmoid}(z)=1/(1+e^{-z})$. At $\beta=0$ this returns $1/(1+e^{-\alpha})$ at every site, the independent case from the end of Section~\ref{sec:model-lattice}. As $\beta S_i$ grows, a site surrounded by occupied neighbors becomes correspondingly unlikely to be occupied itself.

Everything a site needs to know about the rest of the lattice is therefore compressed into the single number $S_i$. The kernel gives every other site a nonzero weight, so $S_i$ formally depends on the whole plane, but the weights $K_{ij}=\exp[-(\|j-i\|_2/\sigma)^p]$ fall off fast enough that the sites beyond any radius $R$ contribute a total that shrinks as $R$ grows. Dropping them perturbs $S_i$ by that amount, and, since $\operatorname{Sigmoid}$ is $1/4$-Lipschitz, perturbs the conditional probability in Eq.~\eqref{eq:local-conditional} by at most $\beta/4$ times that amount. Section~\ref{sec:sampler-truncations} makes this truncation explicit; the supplementary material bounds this discarded mass with an explicit tail-mass estimate for the truncated kernel. What remains is a Markov random field of bounded degree: the conditional law at $i$ involves only the retained neighbors.

\subsection{What the Parameters Control}
\label{sec:model-parameters}

Every parameter of the model changes a statistic of the distribution rather than a property of any single pattern. The statistics that matter for blue noise are the density, the pair correlation, and the power spectrum. Writing $\widehat{u}(\omega)=\sum_{i\in\Lambda}(v_i-\rho)e^{-2\pi\mathrm{i}\,\omega\cdot i}$ for the Fourier transform of the mean-subtracted occupancy field, the expected power at frequency $\omega$ is
\begin{equation}
    S(\omega)
    =
    \frac{1}{|\Lambda|}
    \mathbb{E}\!\left[|\widehat{u}(\omega)|^2\right],
    \label{eq:expected-power}
\end{equation}
which is defined under any boundary condition. Periodic boundaries make the ensemble translation invariant, and $S(\omega)$ is then the Fourier transform of the autocovariance $C(d)=\mathbb{E}[(v_i-\rho)(v_{i+d}-\rho)]$, which no longer depends on $i$. We use them when measuring spectra in Section~\ref{sec:experiments}, since isotropy is only meaningful for a translation-invariant ensemble. None of these statistics is prescribed; each parameter below moves them in a direction we can state.

\paragraph{Density.}
The activity $\alpha$ controls the expected number of occupied sites.
For fixed $\beta$, $\sigma$, $p$, and $\tau$,
\begin{equation}
    \frac{\partial}{\partial\alpha}
    \mathbb{E}_{\mu_\Lambda^\tau}\!\left[N_\Lambda(v)\right]
    =
    \operatorname{Var}_{\mu_\Lambda^\tau}\!\left[N_\Lambda(v)\right]
    \ge 0 ,
    \label{eq:density-monotonicity}
\end{equation}
so raising $\alpha$ raises the expected count, strictly so unless the count is already fixed almost surely. In practice we calibrate $\alpha$ until $\mathbb{E}[\rho_\Lambda(v)]$ matches the requested density $\rho$. The count is controlled in expectation and not fixed exactly: two patterns drawn at the same $\alpha$ will generally contain different numbers of points.

\paragraph{Repulsion.}
The repulsion strength $\beta$ controls how much the ensemble favors spread-out configurations. For fixed $\alpha$, $\sigma$, $p$, and $\tau$,
\begin{equation}
    \frac{\partial}{\partial\beta}
    \mathbb{E}_{\mu_\Lambda^\tau}\!\left[P_\Lambda^\tau(v)\right]
    =
    -\operatorname{Var}_{\mu_\Lambda^\tau}\!\left[P_\Lambda^\tau(v)\right]
    \le 0 ,
    \label{eq:repulsion-monotonicity}
\end{equation}
so raising $\beta$ lowers the expected pair energy. The same holds when $\alpha$ is adjusted to keep $\mathbb{E}[N_\Lambda]$ fixed, as it is whenever $\beta$ is varied below: the derivative becomes $-\operatorname{Var}[P_\Lambda^\tau]\,\bigl(1-\operatorname{corr}(P_\Lambda^\tau,N_\Lambda)^2\bigr)\le 0$, where $\operatorname{corr}$ is the correlation under $\mu_\Lambda^\tau$ (supplementary material). What this does to the spectrum is less simple than it sounds: raising $\beta$ redistributes spectral power across radial modes and organizes the two-dimensional spectrum, and peak heights are not themselves an order parameter for this process. Section~\ref{sec:experiments} shows this directly.

\paragraph{Interaction scale and hardness.}
The kernel parameters act on the spectrum through the distance at which sites compete. Decreasing $\sigma$ shortens that distance, and increasing $p$ makes the kernel approach hard-core repulsion; both move spectral power toward higher frequency, and the change in the height of the principal peak is a consequence of that movement (Figures~\ref{fig:exp-sigma-ablation} and~\ref{fig:exp-p-ablation}). Neither direction follows from an identity like Eqs.~\eqref{eq:density-monotonicity} and~\eqref{eq:repulsion-monotonicity}; both are measured in Section~\ref{sec:experiments}. Changing $\sigma$ or $p$ also changes the effective strength of the repulsion, so those measurements hold the mean-field strength fixed in order to isolate the geometric effect.

Together these four parameters cover a continuous family of blue-noise spectra. Larger $\beta$ makes high-energy configurations exponentially less likely without ever excluding them, so patterns drawn at one parameter setting still differ from one another. The resulting tradeoff between density, order, and diversity is characterized in Section~\ref{sec:experiments}.

\subsection{The Hard-Core Limit}
\label{sec:model-dt-limit}

Every setting above has finite $\beta$, which is what keeps the patterns stochastic. The boundary of the parameter space is a familiar object. As $\beta$ grows with the activity held so that the occupancy probability of an isolated site tends to a limit $s_i$, every retained pair conflict becomes prohibitive, and a single sweep that visits the sites in random order, starting from the empty configuration, reduces to dart throwing on the lattice: each site proposes a point with probability $s_i$, and the proposal survives only if no neighbor visited earlier already holds one. Letting the lattice spacing vanish recovers the continuum version of this process. Dart throwing is therefore not a separate construction to compare against but an endpoint of the sampler's own approximation hierarchy, with hard exclusion in place of soft repulsion and a single pass from the empty state in place of equilibration; it is not the equilibrium law of the model. Both limits are proved in the supplementary material, first on the finite lattice and then in the continuum.

\subsection{Spatially Varying Density}
\label{sec:adaptive}

Applications such as stippling need the point density to follow an image rather than stay constant. The target is then a density map $\rho_i$, one value per lattice site, giving the probability that site $i$ should be occupied. Nothing about the state space changes; two of the model's parameters become fields.

The activity becomes site-dependent, so the energy reads
\begin{equation}
    H_{\alpha,\beta}^{\Lambda,\tau}(v)
    =
    -\sum_{i\in\Lambda} \alpha_i v_i
    +
    \beta P_\Lambda^\tau(v),
    \label{eq:hamiltonian-adaptive}
\end{equation}
and the conditional probability of Eq.~\eqref{eq:local-conditional} becomes $\operatorname{Sigmoid}(\alpha_i - \beta S_i(v^\tau))$, with $\alpha$ replaced by $\alpha_i$ and nothing else changed.

Varying $\alpha_i$ alone is not enough. The interaction scale $\sigma$ sets the distance over which sites compete, and that distance should match how far apart the points are meant to be. Spreading $n$ points over a unit area leaves them about $n^{-1/2}$ apart, so the spacing implied by a density $\rho_i$ goes as $\rho_i^{-1/2}$, and no single $\sigma$ suits the whole map. Where the target density is high, each point is penalized by many more neighbors than the intended spacing calls for, and pairing up becomes the cheaper option: two points sitting next to each other pay one large penalty between them but escape the many mid-range penalties they would otherwise incur. The density is met, but the pattern shows paired points and gaps instead of even spacing. Where the target density is low, the interaction does not reach the intended spacing at all and the sites behave as if $\beta$ were much smaller, falling back toward the clumping of the independent case. We therefore let the scale follow the target density,
\begin{equation}
    \sigma_i = \min\!\left( \sigma_{\min} \sqrt{\frac{\rho_{\max}}{\rho_i}},
                            \; \sigma_{\max} \right),
    \label{eq:adaptive-sigma}
\end{equation}
which gives each site its own profile $k_i(r) = k_{\sigma_i, p}(r)$. Here $\rho_{\max}=\max_i\rho_i$, and $\sigma_{\min}$ is the interaction scale assigned to the densest part of the current map. The expression is understood with the convention $\sigma_i=\sigma_{\max}$ when $\rho_i=0$; if $\rho_{\max}=0$, the target is empty and the scale field is unused. For any fixed nonempty input, $\sigma_{\min}\sqrt{\rho_{\max}}$ is constant, so before the cap is reached Eq.~\eqref{eq:adaptive-sigma} makes $\sigma_i$ proportional to $\rho_i^{-1/2}$. The scale therefore follows the relative variation of the target spacing within the map, anchored at $\sigma_{\min}$ in its densest region. The input map is rescaled so that its maximum equals a fixed value $\rho_{\max}$, so the anchor does not depend on the absolute intensity of the input. The cap at $\sigma_{\max}$ bounds the interaction range in sparse regions.

The energy sums once over each unordered pair, so a pair needs a single weight. When $i$ and $j$ have different scales, $k_i(j-i)$ and $k_j(i-j)$ disagree, and we take
\begin{equation}
    K_{ij} = \min\big(k_i(j-i),\, k_j(i-j)\big) = K_{ji}.
    \label{eq:adaptive-symmetric-weight}
\end{equation}

This defines the adaptive ensemble. What it does not give is the activity field itself: a site's occupancy probability depends on whether its neighbors are occupied, and those neighbors have their own activities, so $\alpha_i$ cannot be read off from $\rho_i$ in closed form. Section~\ref{sec:sampler-adaptive} calibrates it.

\section{A Backward Sampler with Bounded Dependencies}
\label{sec:sampler}

The model in Section~\ref{sec:model} defines a Gibbs ensemble through single-site conditional probabilities. A direct implementation would run a forward Gibbs sampler until the chain is believed to have mixed. For the strongly repulsive parameter regimes that produce blue-noise structure, this stopping criterion is problematic: slow global modes may remain difficult to diagnose even when the local spacing statistics of interest have largely stabilized, so the cost of producing a pattern is not known before it is produced.

To circumvent this ambiguity, we adopt the backward dependency viewpoint of Coupling Towards The Past~\cite{Feng2023Derandomising,liu2026local}: rather than simulating states forward from an arbitrary initialization, we trace the random choices that determine the present state backward through the dependency graph, and construct only the finite portion of that history which is allowed to influence the requested output.

\begin{figure*}[t]
  \centering
  \includegraphics{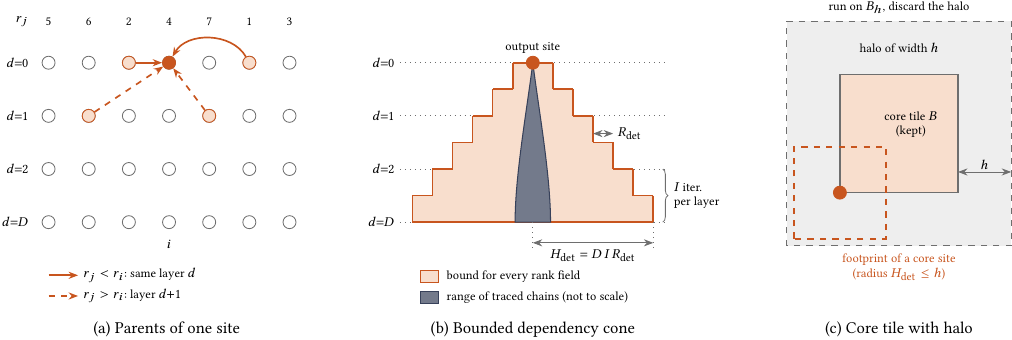}
  \caption{The backward dependency cone, drawn with $D=3$ layers and $I=2$ Jacobi iterations per layer.
  (a)~A one-dimensional slice of the spacetime DAG. The value of site $i$ in layer $d$ ($d=0$ is the output) is read from lower-rank neighbors in the same layer and from higher-rank neighbors in the next deeper layer; the numbers are the ranks $r_j$, and arrows point in the direction of evaluation.
  (b)~Following these edges backward from one output site gives a bounded cone. One Jacobi iteration extends it by at most $R_{\det}$, so after $D$ layers of $I$ iterations its radius is at most $H_{\det}=D\,I\,R_{\det}$, a bound that holds for every rank field (shaded). In practice, chains traced through the read relation (Algorithm~\ref{alg:noisy-dependency-trace}) end far inside this bound (dark region, not to scale): in the configuration of Table~\ref{tab:exp-halo-comparison}, $H_{\det}=600$, while the longest of $67$M traced chains leaves its tile by $39$ pixels (Table~\ref{tab:exp-halo-comparison}).
  (c)~A core tile $B$ generated together with a halo of width $h\ge H_{\det}$ agrees exactly with any larger run that uses the same parameters and coordinate-based random numbers.}
  \label{fig:dependency-cone}
\end{figure*}

The section is organized around this dependency structure. The goal is to make each output site's computational dependencies occupy a bounded spatial region known before sampling begins, so that a tile can be sampled identically to the corresponding region of any larger domain. We first unroll a rank-ordered Gibbs chain into a spacetime directed acyclic graph (DAG). This unrolling is an exact algebraic view of the Markov chain. We then introduce three algorithmic truncations, in spatial range, backward depth, and intra-layer resolution, that turn the unrolled process into a bounded local computation. The resulting finite DAG is evaluated bottom-up by dynamic programming, implemented with Jacobi-style parallel updates, and applied to halo-augmented tiles. The supplementary material proves the properties stated below: the unrolled graph is acyclic and reproduces the chain it unrolls, that relaxing any one truncation recovers the law that truncation approximates, and the error bounds these truncations obey. Figure~\ref{fig:dependency-cone} summarizes this structure.

\subsection{Unrolling the Markov Chain: The Spacetime DAG}
\label{sec:sampler-dag}

Let $V$ denote the finite working domain on which the sampler is executed. In full-image generation $V=\Lambda$; in tiled generation $V$ is a tile enlarged by a halo. For the moment, ignore the finite-range approximation introduced in Section~\ref{sec:sampler-truncations}, and consider a rank-ordered Gibbs sweep on the sites of $V$. We draw a continuous rank field
\begin{equation}
    r_i \sim \mathrm{Uniform}(0,1),
    \qquad i\in V,
\end{equation}
which almost surely gives a strict ordering of the sites. A single sweep visits sites in increasing rank. When site $i$ is updated, all lower-rank sites have already been updated in the current sweep, while higher-rank sites still retain their values from the previous sweep.

We index backward time by $d$, with $d=0$ denoting the output layer and larger $d$ denoting increasingly remote past layers. We also draw an independent coin field
\begin{equation}
    \omega_i^{(d)} \sim \mathrm{Uniform}(0,1),
    \qquad (i,d)\in V\times\{0,\ldots,D-1\}.
\end{equation}
Given the rank and coin fields, the random Gibbs update becomes a deterministic Boolean rule. Drawing all of this randomness in advance is what makes the construction useful: the output is then a deterministic function of these fields, so the set of inputs that a given output site depends on is well defined and can be bounded. For a neighbor $j$ of $i$, the value read by the update at layer $d$ is
\begin{equation}
    \widetilde v_j^{(d)}
    =
    \begin{cases}
        v_j^{(d)}, & r_j < r_i,\\
        v_j^{(d+1)}, & r_j > r_i .
    \end{cases}
    \label{eq:sampler-rank-read}
\end{equation}
Ties occur with probability zero.

Thus each spacetime variable $(i,d)$ depends on lower-rank spatial neighbors within the same layer and higher-rank spatial neighbors in the next deeper layer (Figure~\ref{fig:dependency-cone}a). Writing $\mathcal{N}(i)$ for the represented interacting neighbors of $i$, the parent set is\footnote{Notation: $\mathbf{1}[P]$ denotes the indicator of a condition, equal to $1$ when $P$ holds and $0$ otherwise.}
\begin{equation}
    \partial(i,d)
    =
    \big\{(j, d + \mathbf{1}[r_j>r_i])
      \;:\; j\in\mathcal{N}(i)\cap V\big\} .
    \label{eq:sampler-parent-set}
\end{equation}
The state is then evaluated by the local conditional probability from Eq.~\eqref{eq:local-conditional}:
\begin{equation}
    v_i^{(d)}
    =
    \mathbf{1}
    \left[
        \omega_i^{(d)}
        <
        \operatorname{Sigmoid}
        \left(
            \alpha_i
            -
            \beta
            \sum_{j\in\mathcal{N}(i)} K_{ij}
            v_j^{(d+\mathbf{1}[r_j>r_i])}
        \right)
    \right].
    \label{eq:sampler-dag-update}
\end{equation}
For the homogeneous model, $\alpha_i=\alpha$. For the adaptive model of Section~\ref{sec:adaptive}, $\alpha_i$ is spatially varying and the symmetric weights $K_{ij}$ are read from the adaptive interaction map.

The parent sets define a spacetime graph $\mathcal{G}=(\mathcal{V},\mathcal{E})$, with $\mathcal{V}=V\times\{0,\ldots,D\}$. Equation~\eqref{eq:sampler-dag-update} is the deterministic Boolean update on this graph. Same-layer edges always point from lower rank to higher rank, and inter-layer edges always point from $d+1$ to $d$. Therefore there are no directed cycles.

The unrolled graph is simply another view of the rank-ordered Markov chain: evaluating layers from a remote starting layer down to $d=0$ is equivalent to applying a sequence of Gibbs sweeps from that layer's state. The next subsection makes this construction finite and local.

\subsection{Bounding the Dependency Cone: Space, Depth, and Resolution Truncations}
\label{sec:sampler-truncations}

The spacetime unrolling above is lossless for the chosen Markov chain, but it is not yet a practical graphics algorithm. Without additional restrictions, a query may have an unbounded spatial fan-in, an unbounded temporal history, and, under exact rank-ordered evaluation, long same-layer dependency chains. We make three truncations explicit.

\paragraph{Spatial truncation.}

The ideal kernel in Section~\ref{sec:model} has infinite support. In the sampler we replace it by a finite, translation-invariant retained edge set $E_{\mathrm{tr}}\subset\{\{i,j\}: i,j\in\mathbb{Z}^2,\ i\ne j\}$, where membership depends only on the offset $j-i$, and set
\begin{equation}
    K^{\mathrm{tr}}_{ij}
    =
    k_{\sigma,p}(j-i)\,
    \mathbf{1}[\{i,j\}\in E_{\mathrm{tr}}].
    \label{eq:sampler-truncated-kernel}
\end{equation}
The corresponding finite neighborhood and maximum degree are
\begin{equation}
    \mathcal{N}_{\mathrm{tr}}(i)
    =
    \{j\in\mathbb{Z}^2:\{i,j\}\in E_{\mathrm{tr}}\},
    \qquad
    M_{\mathrm{tr}}
    =
    \max_{i\in V}|\mathcal{N}_{\mathrm{tr}}(i)|.
    \label{eq:sampler-neighborhood}
\end{equation}
A retained neighbor of $i\in V$ need not itself lie in $V$; when it does not, its value is supplied by the boundary convention of Section~\ref{sec:model} rather than by discarding the edge.

The retained edge set can be specified in either of two modes. In explicit radius mode, we retain all lattice offsets whose Euclidean distance is at most the supplied radius:
\begin{equation}
    \{i,j\}\in E_{\mathrm{tr}}
    \quad\Longleftrightarrow\quad
    0<\|i-j\|_2\le r_{\mathrm{explicit}} .
    \label{eq:sampler-explicit-radius}
\end{equation}
In implicit threshold mode, we retain exactly those offsets whose kernel weight exceeds a fixed tolerance $\epsilon$:
\begin{equation}
    \{i,j\}\in E_{\mathrm{tr}}
    \quad\Longleftrightarrow\quad
    k_{\sigma,p}(j-i)>\epsilon .
    \label{eq:sampler-threshold-neighborhood}
\end{equation}
For the generalized-Gaussian kernel, this threshold corresponds to the Euclidean radius
\begin{equation}
    r_{\mathrm{implicit}}
    =
    \sigma\,\bigl(-\log\epsilon\bigr)^{1/p}.
    \label{eq:sampler-implicit-radius}
\end{equation}
We use $\epsilon=10^{-6}$ in our implementation. The implicit and explicit modes are not combined; when an explicit radius is supplied, it takes precedence.

For deterministic locality bounds, we use the conservative lattice radius
\begin{equation}
    R_{\det}
    =
    \max_{\{i,j\}\in E_{\mathrm{tr}}}
    \|i-j\|_\infty .
    \label{eq:sampler-deterministic-radius}
\end{equation}
This radius is used only for the dependency-cone bound; it does not enlarge the retained edge set and therefore does not change the truncated Gibbs target. Thus the actual retained stencil may be a Euclidean disk or a weight-threshold set, while the halo analysis uses its $\ell_\infty$ bounding radius
$R_{\det}$.

This produces a finite-range target Gibbs measure, denoted $\pi_{\mathrm{tr}}$. Let $\pi$ denote the ideal full-kernel Gibbs measure of Section~\ref{sec:model} on the current working domain $V$, under the chosen boundary convention and model parameters. The analysis in supplementary material bounds the discarded mass and treats the difference between the ideal full-kernel measure $\pi$ and the retained measure $\pi_{\mathrm{tr}}$ as a spatial tail error. In the remainder of this section, $K_{ij}$ and $\mathcal{N}(i)$ refer to the retained weights $K^{\mathrm{tr}}_{ij}$ and the finite neighborhood $\mathcal{N}_{\mathrm{tr}}(i)$ unless otherwise stated.

\paragraph{Depth truncation.}
A truly stationary backward construction would require an infinite past. We instead keep only $D$ backward layers and initialize the deepest layer from a simple product distribution $q$:
\begin{equation}
    v_i^{(D)} \sim \mathrm{Bernoulli}(q_i),
    \qquad i\in V .
    \label{eq:sampler-boundary-layer}
\end{equation}
For uniform examples we use a constant $q_i$ matched to the target density; for adaptive examples we use the local density map. The exact evaluation of this finite-depth DAG defines the approximation $\hat\pi_{\mathrm{tr},D}$. The parameter $D$ is not a thermodynamic temperature. It controls how much backward dependency history is retained. Large $D$ approaches the finite-range retained equilibrium measure $\pi_{\mathrm{tr}}$. The supplementary material treats the difference between $\hat\pi_{\mathrm{tr},D}$ and $\pi_{\mathrm{tr}}$ as a finite-depth error and bounds it in terms of $D$.

\paragraph{Intra-layer resolution truncation.}
Given $v^{(d+1)}$, the variables $v^{(d)}$ form an acyclic Boolean system because same-layer dependencies follow increasing rank. A sequential reference implementation could evaluate the sites once in rank order. On parallel hardware, however, we solve the same layer by Jacobi-style iterations. Each iteration propagates information across one same-layer dependency edge. If the iteration is run until the Boolean state no longer changes, the exact finite-depth DAG value is recovered. In practice we cap the number of Jacobi iterations by $I$, giving the operational distribution $\hat\pi_{\mathrm{tr},D,I}$. The supplementary material treats the difference between $\hat\pi_{\mathrm{tr},D,I}$ and $\hat\pi_{\mathrm{tr},D}$ as a Jacobi-resolution error and bounds it in terms of $I$.

\subsection{Bottom-Up Parallel Reconstruction: The Dynamic Programming View}
\label{sec:sampler-dp}

A naive backward sampler would recursively expand dependencies for every queried site. Neighboring queries would repeatedly visit the same spacetime nodes, causing substantial redundant work and irregular memory access. The DAG view suggests the opposite evaluation order: compute complete layers from the bottom up, reuse each resolved layer as the boundary condition for the next one, and discard layers that are no longer needed.

For a fixed rank field, coin field, and deepest layer $v^{(D)}$, the finite-depth pattern is deterministic. Algorithm~\ref{alg:cttp_sampler} shows the operational reconstruction. The notation $x$ denotes the current Jacobi iterate for layer $d$. Lower-rank dependencies read from the previous Jacobi iterate $x^{\mathrm{old}}$, while higher-rank dependencies read from the already-known deeper layer $v^{(d+1)}$. If the Boolean state reaches a fixed point before the iteration budget is exhausted, the layer is exactly resolved. Otherwise the capped iterate defines the intended approximation $\hat\pi_{\mathrm{tr},D,I}$.

\begin{algorithm}[t]
\caption{Parallel DAG-based Backward Sampler}
\label{alg:cttp_sampler}
\begin{algorithmic}[1]
\Require Working domain $V$, retained graph $E_{\mathrm{tr}}$ with weights
$K^{\mathrm{tr}}$, activity field $\{\alpha_i\}$, repulsion $\beta$, depth $D$,
Jacobi budget $I$, deepest-layer probabilities $\{q_i\}$, boundary convention
$\tau$
\Ensure Occupancy pattern $v^{(0)}$ on $V$
\State Draw ranks $r_i\sim\mathrm{Uniform}(0,1)$ for all $i\in V$
\State Draw coins $\omega_i^{(d)}\sim\mathrm{Uniform}(0,1)$ for all $i\in V$ and $d=0,\ldots,D-1$
\State Draw $v_i^{(D)}\sim\mathrm{Bernoulli}(q_i)$ for all $i\in V$
\For{$d=D-1$ \textbf{down to} $0$}
    \State $x_i \gets v_i^{(d+1)}$ for all $i\in V$
    \For{$t=1$ \textbf{to} $I$}
        \State $x^{\mathrm{old}} \gets x$
        \ParallelFor{each site $i\in V$}
            \State $S_i \gets \displaystyle\sum_{j\in\mathcal{N}_{\mathrm{tr}}(i)}
            K^{\mathrm{tr}}_{ij}
            \begin{cases}
                x^{\mathrm{old}}_j, & j\in V,\ r_j<r_i,\\
                v_j^{(d+1)}, & j\in V,\ r_j>r_i,\\
                \tau_j, & j\notin V
            \end{cases}$
            \State $x_i \gets \mathbf{1}\!\left[\omega_i^{(d)} < \operatorname{Sigmoid}\left(\alpha_i-\beta S_i\right)\right]$
        \EndFor
        \If{$x=x^{\mathrm{old}}$}
            \State \textbf{break}
        \EndIf
    \EndFor
    \State $v^{(d)}\gets x$
\EndFor
\State \Return $v^{(0)}$
\end{algorithmic}
\end{algorithm}

The third branch applies only when a retained neighbor carries no rank, which happens under a vacuum or randomized exterior but not under periodic wrapping, where the neighbor maps back into $V$. The value $\tau_j$ must depend only on the site, since two tiles whose halos both cover $j$ have to read the same value there.

By replacing the global asymptotic convergence of a traditional Markov chain with a dynamically capped backward evaluation, the runtime of our sampler becomes strictly deterministic. The loop structure in Algorithm~\ref{alg:cttp_sampler} gives a worst-case work bound of $O(D I M_{\mathrm{tr}}|V|)$. There are $D$ layers, at most $I$ Jacobi sweeps per layer, and each sweep evaluates an $M_{\mathrm{tr}}$-neighbor retained stencil. Unlike forward MCMC methods where the required number of sweeps for strongly repulsive regimes is highly variable and difficult to diagnose, our cost is rigorously bounded a priori.

Furthermore, the active storage remains highly localized. Assuming ranks and coins are generated by coordinate-based counter random numbers (e.g., Philox), the auxiliary random fields do not need to be materialized. The working memory is therefore $O(|V|)$ to hold the current iterate, previous iterate, and the deeper layer. We formalize these bounds, together with the operational-locality proof, in the supplementary material.

\subsection{Bounded Locality and Halo-Based Tiled Generation}
\label{sec:sampler-halo}

The bounded dependency cone makes the sampler tileable. Let $B$ be the core
tile whose output will be kept, and let
\begin{equation}
    B_h
    =
    \{i\in\mathbb{Z}^2 : \mathrm{dist}_\infty(i,B)\le h\}.
    \label{eq:sampler-halo-tile}
\end{equation}
be the tile enlarged by a halo of width $h$. We run
Algorithm~\ref{alg:cttp_sampler} on $B_h$ and discard the halo output.

A crucial property of our operational solver with retained graph $E_{\mathrm{tr}}$ and parameters $(D,I)$ is that its computational dependencies form a strictly bounded spacetime cone (Figure~\ref{fig:dependency-cone}b). Information can propagate across at most one retained edge during a single Jacobi iteration. Since every retained edge has $\ell_\infty$ length at most $R_{\det}$, one Jacobi iteration can expand the dependency set by at most $R_{\det}$ in the $\ell_\infty$ metric. Accumulated over $I$ intra-layer iterations and $D$ backward layers, the algorithm possesses the deterministic conservative cone radius
\begin{equation}
    H_{\mathrm{det}}(E_{\mathrm{tr}},D,I)
    =
    D I R_{\det}.
    \label{eq:sampler-deterministic-cone}
\end{equation}

The supplementary material's operational-locality proposition provides the formal proof of this property. The operational consequence is: if the halo width $h\ge H_{\mathrm{det}}(E_{\mathrm{tr}},D,I)$, every core value in $B$ is identical to the value obtained by running the exact same solver on any larger domain containing $B_h$, provided the local random variables are generated from consistent coordinate-based seeds (Figure~\ref{fig:dependency-cone}c).

While $H_{\mathrm{det}}$ guarantees exact mathematical equivalence, it is intentionally conservative. In practice, increasing the retained stencil, $D$, $I$, and $h$ trades computation for fidelity, and the visually relevant boundary influence often decays much faster than the worst-case cone radius. The role of this subsection is only to describe the implementation mechanism: halo generation is possible because the operational computation is local by construction. In the supplementary material, we separate the corresponding errors into spatial tail error, finite-depth error, Jacobi-resolution error, and boundary or halo error, following the local error-decomposition inequality stated at the start of the theoretical analysis.

\subsection{Adaptive Generation and Local Density Approximation}
\label{sec:sampler-adaptive}

The same DAG reconstruction applies to the spatially varying model in Section~\ref{sec:adaptive}. The update rule is unchanged except that the activity $\alpha_i$ and interaction weights $K_{ij}$ are read from spatial parameter maps. For a target density field $\rho_i$, we use the adaptive scale from Section~\ref{sec:adaptive} and construct a symmetric retained master graph $E_{\ast}$. The master graph contains every pair that can have a nonzero retained interaction under the allowed range of local scales. For example, in implicit threshold mode it can be constructed from the largest allowed scale $\sigma_{\max}$ and the same tolerance $\epsilon$ used in the homogeneous sampler; in explicit radius mode it is constructed from the supplied maximum Euclidean radius.

For each retained edge, we use the symmetric weight
\begin{equation}
    K_{ij}
    =
    \min\!\left(k_{\sigma_i,p}(j-i), k_{\sigma_j,p}(i-j)\right)
    \mathbf{1}[\{i,j\}\in E_{\ast}],
    \label{eq:sampler-adaptive-weight}
\end{equation}
and the corresponding adaptive neighborhood is
\begin{equation}
    \mathcal{N}_\ast(i)
    =
    \{j:\{i,j\}\in E_{\ast}\}.
\end{equation}
The deterministic lattice radius used by the halo bound is
\begin{equation}
    R_{\det,\ast}
    =
    \max_{\{i,j\}\in E_{\ast}}\|i-j\|_\infty .
\end{equation}
This keeps the dependency cone bounded by the same form as Eq.~\eqref{eq:sampler-deterministic-cone}, with $E_{\mathrm{tr}}$ and $R_{\det}$ replaced by $E_{\ast}$ and $R_{\det,\ast}$.

The remaining adaptive problem is calibration: we need an activity field $\alpha_i$ whose local equilibrium density matches $\rho_i$. The mapping from activity to density is monotone but nonlinear, so we use a local density approximation (LDA). For fixed $(\beta,p,\sigma_{\min},\sigma_{\max})$, we precompute a lookup table $A(\rho)$ on homogeneous periodic calibration patches. For each density bin, the scale $\sigma(\rho)$ is set by the adaptive scale rule, and a one-dimensional stochastic root find adjusts $\alpha$ until the empirical density of the sampler matches the bin density.

During generation, we set
\begin{equation}
    \alpha_i = A(\rho_i).
    \label{eq:sampler-lda-alpha}
\end{equation}
by linear interpolation in the table, compute $\sigma_i$ and $K_{ij}$ locally, and then run the same parallel DAG reconstruction as in Algorithm~\ref{alg:cttp_sampler}. The LDA assumes that the density field varies slowly relative to the local exclusion scale; under that assumption, each neighborhood behaves approximately like a homogeneous Gibbs ensemble with the same local density and scale. Rapidly varying density maps can still be processed by the algorithm, but the density-matching error is then an LDA calibration error rather than a failure of the DAG reconstruction itself.

\section{Experiments}
\label{sec:experiments}

\subsection{Experimental Protocol}
\label{sec:exp-setup}

\begin{figure*}[t]
  \centering
  \includegraphics{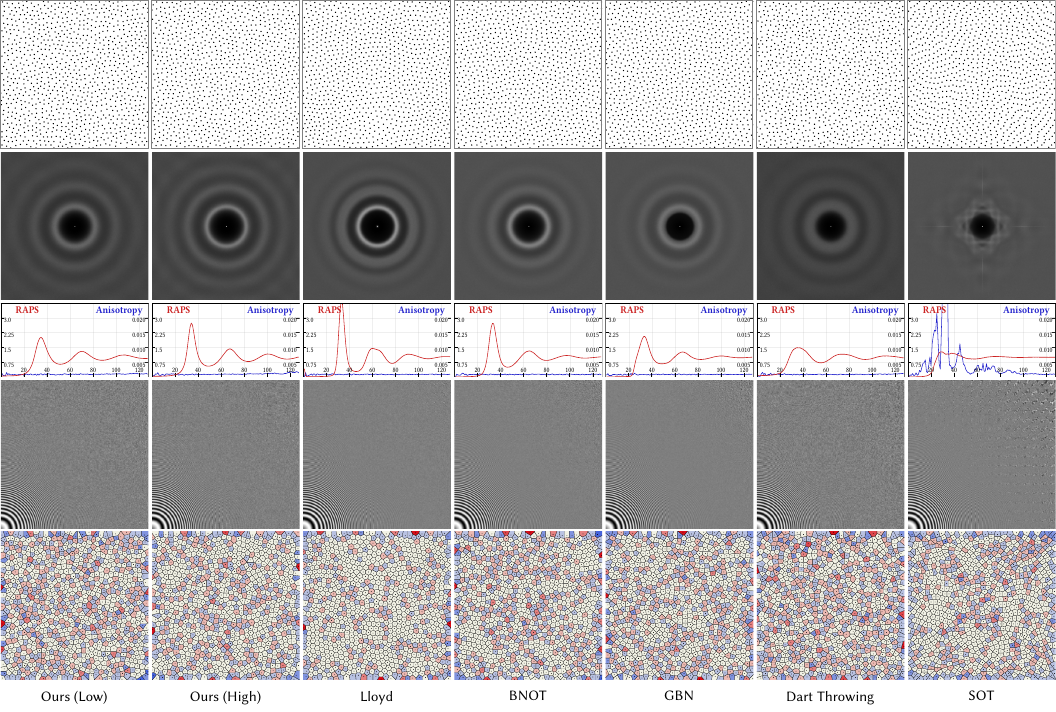}
    \caption{\textbf{Homogeneous blue-noise comparison against reference methods.} Matched density ($N = 983$, $\rho_0 = 983/256^2 \approx 0.0150$, $256\times256$ periodic domain), averaged over $1000$ independent runs. Each column is one configuration, including two operating points of our sampler at $D=12$ (Ours(Low)) and $D=128$ (Ours(High)); rows show (1)~spatial patterns, (2)~power spectral density (PSD), (3)~radially averaged power spectrum (RAPS) with the anisotropy curve below it, (4)~zone-plate response, and (5)~Delaunay valence distributions. All seven configurations produce well-formed low-frequency voids and principal rings, placing the proposed ensemble in the standard blue-noise regime. Anisotropy is well controlled everywhere except SOT, whose axis-aligned spectral energy also appears as a visible alias in the zone-plate row beyond the Nyquist limit. Differences in that frequency range lie outside the target band of blue-noise sampling and reflect each method's intrinsic sampling character rather than a deficiency.}
  \label{fig:exp-homogeneous-reference}
\end{figure*}

All experiments were run on a laptop workstation equipped with an NVIDIA RTX 5070Ti Laptop GPU and an Intel Core Ultra 9 285H CPU, except BNOT and SOT, which require platform-specific dependencies and were run on an Apple MacBook Air (M4, 10-core CPU and 10-core GPU). Because these two run on different hardware, we report no timing comparison across methods; the timings given in this section are absolute costs for our own configurations. Homogeneous experiments use periodic boundary conditions. Plotted spectra are normalized by the number of points in each pattern rather than by $|\Lambda|$ as in Eq.~\eqref{eq:expected-power}, so that uncorrelated points have power close to $1$.

The model's four native parameters $(\alpha, \beta, \sigma, p)$ determine the Gibbs measure, and the density follows from them; since density is what a user wants to set directly, we invert this relation, which costs one of the four its freedom. We give up $\alpha$, which alone among the four carries no spectral role, as the others set repulsion strength and kernel geometry.  $\alpha$ is calibrated once per configuration against the target density $\rho_0$ with that configuration's own sampler, then held fixed while that configuration's evaluation patterns are generated. For the model itself, $\alpha$ is uniquely determined by the density constraint; for the sampler of a given configuration the supplementary material proves that a solution exists, and we find one numerically with the procedure given there. All timings reported in this section are for pattern generation.\footnote{Calibration is not included: $\alpha$ is an input to the sampler rather than a byproduct of running it. Our implementation accepts $\alpha$ directly via a command-line flag, bypassing calibration entirely, and generation proceeds identically regardless of how $\alpha$ was obtained.}

Across the comparative and ablation experiments the domain is a $256 \times 256$ periodic lattice at target density $\rho_0 = 0.015$, and parameters not under study are held at $\sigma = 6.2$, $p = 6$, and $\beta = 20$;\footnote{$p=6$ and $\beta=20$ are unremarkable choices within a broad range that avoids two failure modes: $p$ too small paired with $\sigma$ too large compounds a slowly decaying tail with a large physical scale, inflating the retained neighborhood well beyond what a well-behaved kernel needs and eroding the locality this sampler otherwise guarantees, and $\beta$ too low approaches a regime where repulsion is too weak to be of interest. Neither is tuned further. Given these, $\sigma=6.2$ satisfies two constraints specific to this section: most reference methods exhibit three principal rings, and a substantially different $\sigma$ shifts this count enough to hinder comparison; within that range, $\sigma \le 6$ leaves too little separation between $D=12$ and $D=128$ to illustrate the two operating points introduced in Section~\ref{sec:exp-blue-noise}.} unless stated otherwise, an experiment departs from these only for the parameter it sweeps. Two conventions recur. First, unless an experiment sweeps the truncation radius, the retained neighborhood is chosen implicitly, keeping an offset when its kernel weight exceeds $10^{-6}$. Second, in experiments that vary the kernel we hold the mean-field interaction strength $\beta \cdot f_{\mathrm{pack}}$ constant, where
\begin{equation}
  f_{\mathrm{pack}} = \rho_0\,A(\sigma,p,R),
  \qquad
  A(\sigma,p,R) = \sum_{\delta \in \mathcal{N}(R)}
  \exp\!\left[-\left(\frac{\|\delta\|}{\sigma}\right)^p\right],
\end{equation}
so that spectral changes are attributable to kernel geometry rather than to a change in effective repulsion. The full configuration of every homogeneous experiment is listed in the supplementary material.

\subsection{Homogeneous Blue-Noise Statistics and Reference Comparisons}
\label{sec:exp-blue-noise}

We first evaluate the homogeneous setting, where the target density is constant over a periodic square domain. To illustrate performance across different computational budgets, we include two operating points of our sampler under the same kernel and $\beta$ configuration: $I$ is run to its fixed point, $R$ is set to the automatic threshold ($\epsilon = 10^{-6}$), and we vary only $D$, showing $D=12$ (Ours(Low)) and $D=128$ (Ours(High)). Since the layer solve is run to its fixed point and the automatic threshold retains over $99.99\%$ of the kernel mass, depth $D$ is the only truncation left to vary, which reduces the budget comparison to a single axis. Generating one $256\times256$ pattern takes $13.0$~ms for Ours(Low) and $155.6$~ms for Ours(High), averaged over the $1000$ runs. These comparisons are statistical reference anchors rather than a ranking under a single objective: each method optimizes something different. What distinguishes our method is not spectral quality alone but that it achieves competitive quality while remaining local and supporting tiled execution at arbitrary scale, properties none of the reference methods provide. Tile-based methods, which do offer locality, are not among the reference methods: their output has the spectral quality of the offline generator that produced the tiles, so a spectral comparison would evaluate that generator rather than the tiling (Section~\ref{sec:related}).

\begin{figure*}[t]
  \centering
  \includegraphics{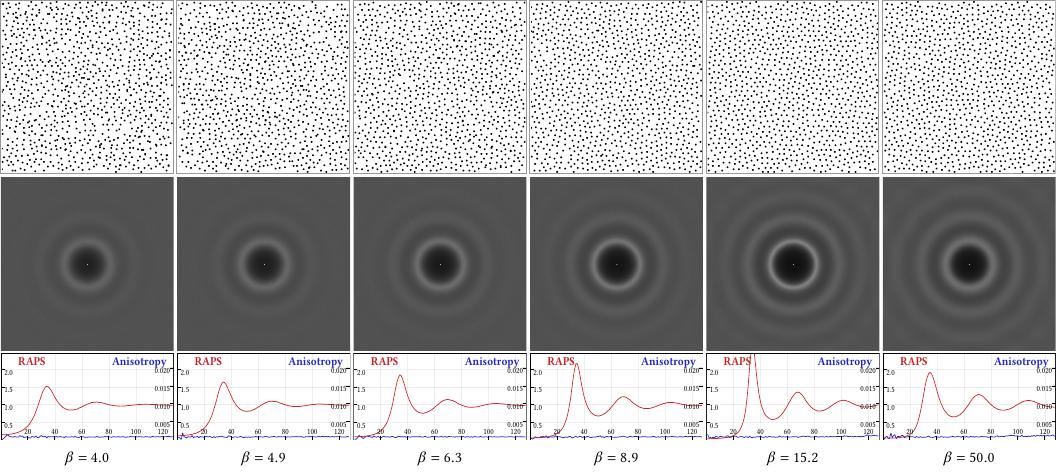}
    \caption{\textbf{Repulsion-strength ablation.} Fixed kernel $(\sigma=6.2, p=6)$ and density $\rho_0=0.015$; each column is one $\beta$ value. Rows show spatial patterns, power spectral density, and RAPS with anisotropy. As $\beta$ increases, power is pushed out of the low-frequency band and accumulates in the ring structure at higher frequency. The first RAPS peak height is not an order parameter for this redistribution: $\beta=6.3$ and $\beta=50$ reach comparable peak magnitudes under the shared vertical scale, yet their two-dimensional spectra differ markedly.}
  \label{fig:exp-beta-ablation}
\end{figure*}

Reference results come from released implementations, called without modification and given only the target point count. GBN uses the configuration reported in its paper, run from the authors' code ported to our platform. Lloyd relaxation and dart throwing have no released implementation we could use, so we implemented them; each has a single free choice, stated in the supplementary material. We did not tune any method beyond these settings. All statistics are ensemble averages over $1000$ independent runs from randomly drawn initial configurations.

Figure~\ref{fig:exp-homogeneous-reference} summarizes the comparison. All seven configurations produce visually well-distributed spatial patterns with clear low-frequency voids and principal rings in the PSD, confirming that the proposed ensemble lies within the standard blue-noise regime.

Anisotropy is well controlled across every configuration except SOT, whose two-dimensional spectrum carries axis-aligned energy, visible as a cross through the origin; the same structure appears in the SOT results reported by \citet{Ahmed2022GaussianBN}.

The first RAPS peak is highest for Lloyd; Ours(High) and BNOT reach comparable levels, as do Ours(Low) and GBN; dart throwing is second-lowest and SOT lowest. The zone plate shows a similar picture across all seven configurations. SOT alone shows a visible alias beyond the Nyquist limit, which is the image-domain counterpart of its axis-aligned spectral energy rather than a separate deficiency. In the Delaunay valence distributions, Lloyd shows the highest proportion of six-valence cells, followed by SOT; Ours(High), GBN, and BNOT sit at a comparable level; Ours(Low) shows a modestly lower proportion; dart throwing shows the lowest.

\subsection{Thermodynamic Control of Density, Order, and Spectral Shape}
\label{sec:exp-thermodynamic-control}

We next evaluate the parameters of the Gibbs model: the repulsion strength $\beta$ and the generalized Gaussian kernel $(\sigma, p)$.  The aim is to verify that they provide predictable control over the resulting blue-noise ensemble.  In the thermodynamic reading of Section~\ref{sec:model}, $\beta$ is an inverse temperature; we use the term repulsion strength throughout, since that is what it controls in the output.  In both experiments the target density $\rho_0 = 0.015$ is held fixed, and the mean-field interaction strength $\beta \cdot f_\mathrm{pack}$ is kept constant across kernel configurations so that spectral changes are attributable solely to kernel geometry rather than to changes in effective repulsion strength.

\begin{figure}[htpb]
  \centering
  \includegraphics[width=\linewidth]{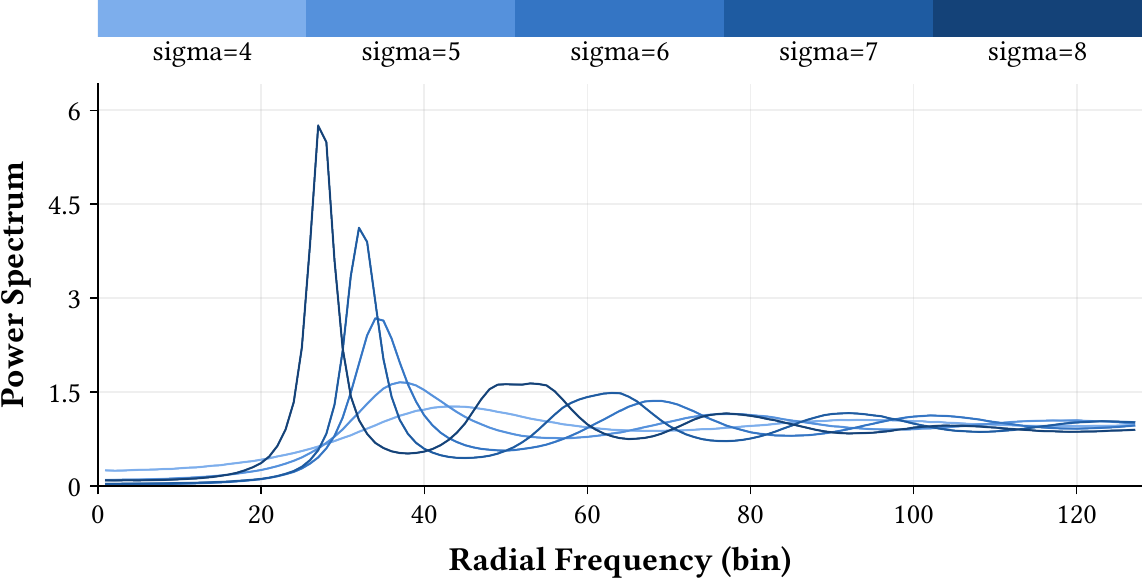}
    \caption{\textbf{Kernel-scale ablation.} Fixed $p=6$, density $\rho_0=0.015$, and mean-field strength $\beta \cdot f_\mathrm{pack}$; each curve is one $\sigma \in \{4,5,6,7,8\}$. Decreasing $\sigma$ moves spectral power to higher frequency, so the principal peak shifts to higher frequency and is lowered, consistent with $\sigma$ setting the characteristic repulsion distance.}
  \label{fig:exp-sigma-ablation}
\end{figure}

\begin{figure}[htpb]
  \centering
  \includegraphics[width=\linewidth]{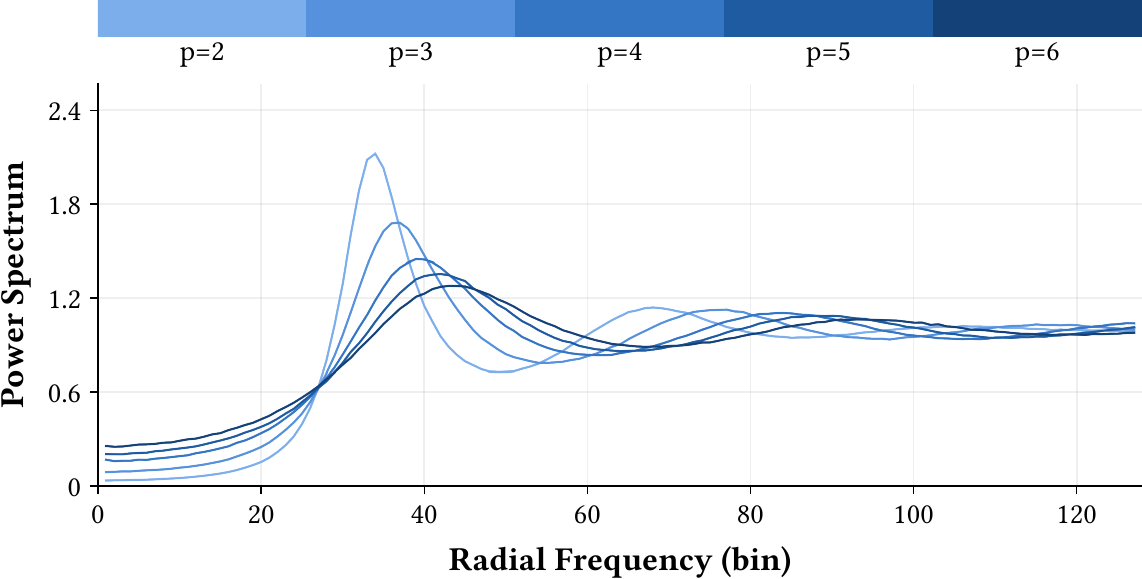}
    \caption{\textbf{Kernel-hardness ablation.} Fixed $\sigma=4$, density $\rho_0=0.015$, and mean-field strength $\beta \cdot f_\mathrm{pack}$; each curve is one $p \in \{2,3,4,5,6\}$. Increasing $p$ moves the kernel toward hard-core repulsion and moves spectral power to higher frequency; the principal peak is lowered as a result.}
  \label{fig:exp-p-ablation}
\end{figure}

\paragraph{Repulsion strength.}
Equation~\eqref{eq:repulsion-monotonicity} and its fixed-density form show that raising $\beta$ lowers the expected pair energy: configurations with nearby occupied pairs are suppressed more strongly. This has two consequences. First, once $\alpha$ is recalibrated to hold the density fixed, the spectral power averaged over frequencies is $\rho_0(1-\rho_0)$ regardless of $\beta$, so suppressing low-frequency fluctuations must move power toward higher frequencies rather than remove it. Second, this redistribution has an endpoint: as $\beta$ grows, repulsion approaches hard exclusion, the regime described in Section~\ref{sec:model-dt-limit}. Figure~\ref{fig:exp-beta-ablation} confirms the first and shows the trend toward the second up to $\beta=50$.

\paragraph{Kernel scale and hardness.}

The kernel $k_{\sigma,p}(r)=\exp[-(r/\sigma)^p]$ separates into a scale and a shape. The penalty falls to $e^{-1}$ at $r=\sigma$ regardless of $p$, so $\sigma$ sets the repulsion distance and therefore the characteristic spacing; the corresponding spectral features sit near $1/\sigma$. The exponent $p$ controls how sharply the penalty falls off: as $p$ grows the kernel approaches a hard disk of radius $\sigma$, tightening the effective interaction range. To isolate these geometric effects, we hold $\beta \cdot f_{\mathrm{pack}}$ constant across kernel configurations.\footnote{$f_{\mathrm{pack}} = \rho_0 A(\sigma, p, R)$, where $A(\sigma, p, R) = \sum_{\delta \in \mathcal{N}(R)} \exp[-(\|\delta\|/\sigma)^p]$ is the kernel mass. The reference value is computed at $\sigma=6.2$, $p=6$, the default configuration of Section~\ref{sec:exp-blue-noise}; any fixed constant is equally valid since the goal is to isolate geometric effects.} Figures~\ref{fig:exp-sigma-ablation} and~\ref{fig:exp-p-ablation} confirm both predictions.

Together, these ablations show that $(\beta,\sigma,p)$ continuously span a family of spectral profiles. Characterizing this family, and the inverse problem of locating the configuration that best approximates a given target spectrum, are left to future work.
\subsection{Sampler Approximation, Locality, and Scaling}
\label{sec:exp-approximation}

This section evaluates the three truncation parameters of the sampler ($D$, $I$, $R$), the practical halo bounds used for tiled execution, and the empirical time complexity.  All experiments in this section use a homogeneous grid with periodic boundary conditions; spectral observables are reported as ensemble averages over independent rank-field seeds.

\paragraph{Backward depth.}

\begin{figure}[htpb]
  \centering
  \includegraphics{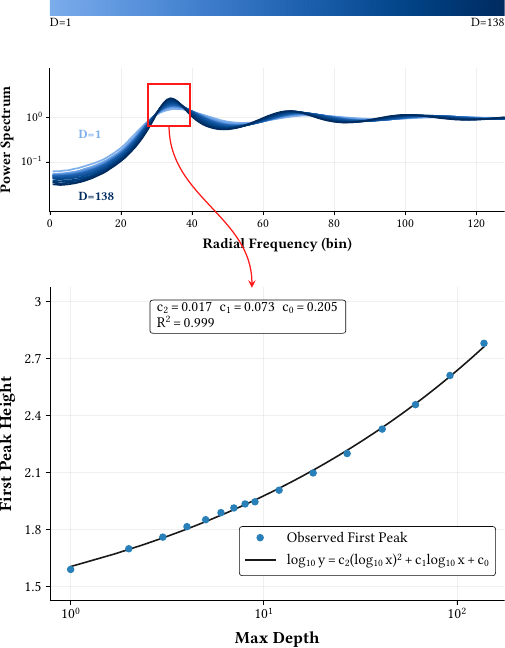}
    \caption{\textbf{Backward-depth ablation.} $\sigma=6.2$, $p=6$, $\beta=20$; $\alpha$ recalibrated per $D$. Top: RAPS overlay for $D \in \{1,\dots,9,12,18,27,41,61,92,138\}$. Bottom: first-peak height versus $D$ on logarithmic axes, with a quadratic fit in log--log space ($R^2=0.999$) whose curvature is positive ($c_2=0.017$): the peak keeps rising over the swept range and shows no sign of saturating.}
  \label{fig:exp-depth-ablation}
\end{figure}

As established in Section~\ref{sec:sampler-dag}, evaluating $D$ backward layers is equivalent to applying $D$ rank-ordered Gibbs sweeps from the initial state. Each additional sweep lets the repulsion structure propagate through one more level of conditional dependence. Figure~\ref{fig:exp-depth-ablation} shows that the first RAPS peak rises monotonically with $D$. The gain per additional sweep shrinks, but the gain per decade of $D$ does not, and the peak is still rising at $D=138$. We therefore read $D=12$ and $D=128$ as two points on a cost--quality curve rather than as converged ensembles. The supplementary material bounds the distance to the finite-range target only by a worst-case rate in $D$ that is not practical to estimate, so we treat $D$ as a fidelity parameter and validate it empirically.

\paragraph{Jacobi iteration count.}

\begin{figure}[htpb]
  \centering
  \includegraphics[width=\linewidth]{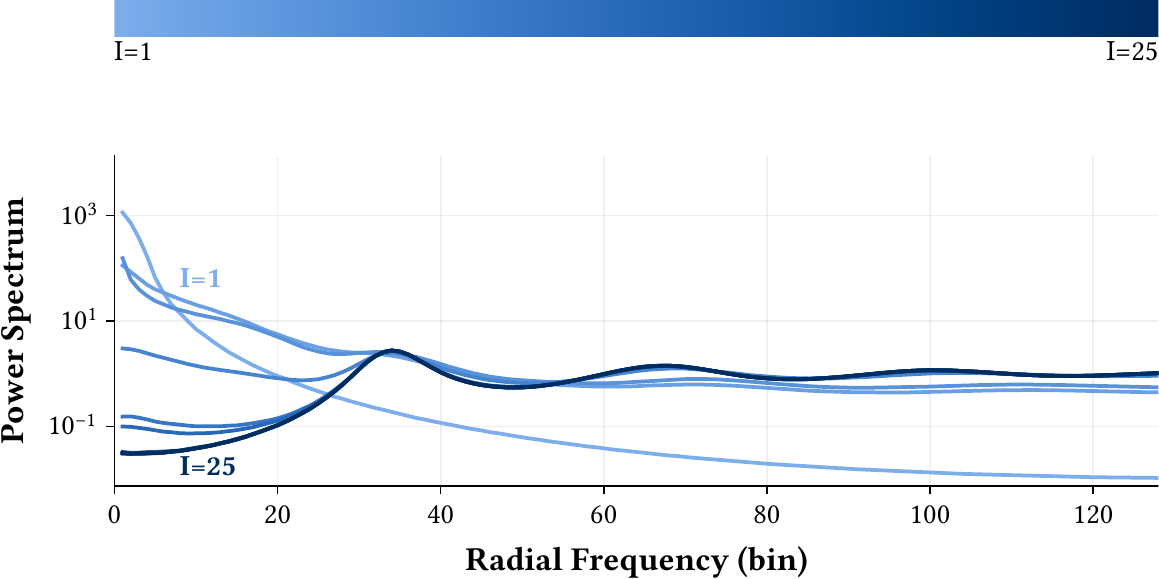}
    \caption{\textbf{Jacobi-iteration ablation.} $\alpha$ fixed at $\alpha^\star$, calibrated by running the Jacobi solve to its fixed point; early-exit test disabled so that each setting runs its full $I$. Low $I$ values fail both spectrally (no principal peak) and in density (deviation from target), confirming that too few iterations leave the layer unresolved. A small misalignment in the low-frequency region remains at $I=7$; for $I \geq 11$ no comparable misalignment is visible, and the curves are close to indistinguishable from $I=25$.}
  \label{fig:exp-iters-ablation}
\end{figure}

Figure~\ref{fig:exp-iters-ablation} shows the RAPS for $I \in \{1,\dots,7\} \cup \{11, 15, 20, 25\}$. Unlike the depth ablation, here we fix $\alpha = \alpha^\star$, calibrated once by running the Jacobi solve to its fixed point (until the layer stops changing, i.e.\ $x=x^{\mathrm{old}}$ in Algorithm~\ref{alg:cttp_sampler}), and hold it constant across all $I$ values. This is because $\alpha$ calibration relies on a stable density response function, which low $I$ values do not provide: at low $I$ the Jacobi solver has not reached a fixed point, the conditional density is not a well-defined function of $\alpha$, and recalibration would produce a fictitious value with no operational meaning. The early-exit test is disabled here so that each setting runs its full $I$.

\paragraph{Spatial truncation radius.}

\begin{figure}[htpb]
  \centering
  \includegraphics[width=\linewidth]{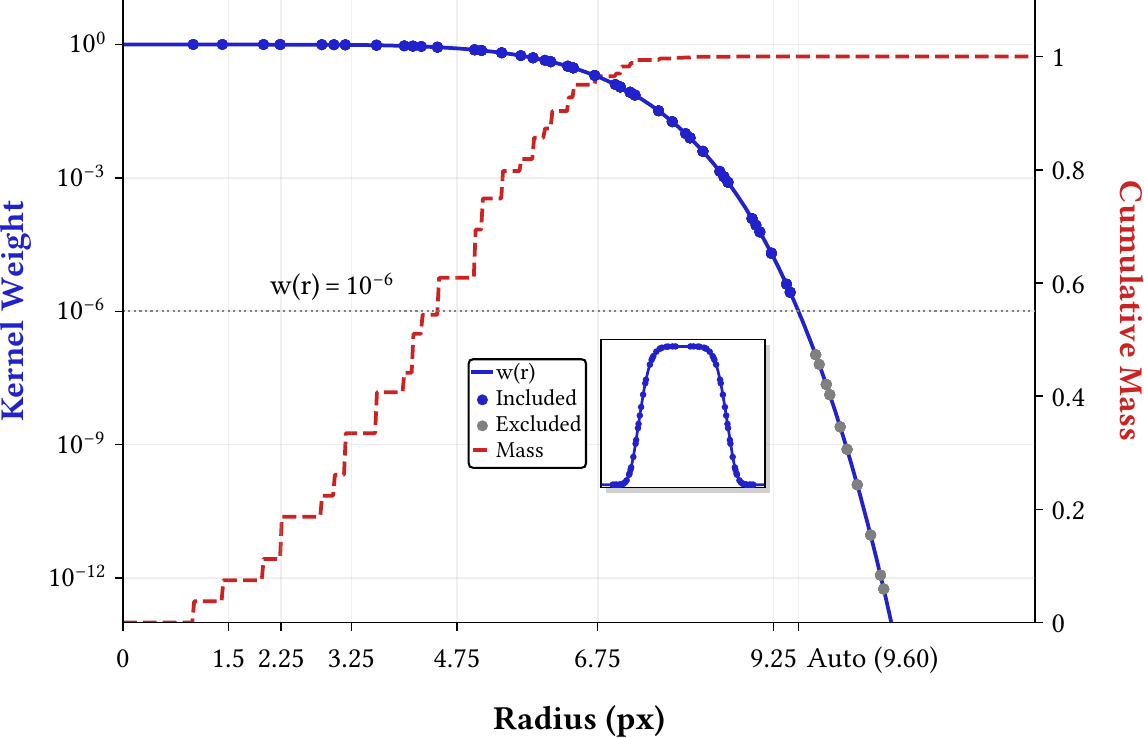}
    \caption{\textbf{Kernel profile and tail.} Main panel: kernel tail on a log axis with cumulative captured mass on the twin axis; shaded bands mark the tested truncation radii. Inset: kernel shape on a linear axis, showing the near-plateau profile of the generalized Gaussian at $p=6$.}
  \label{fig:exp-r-ablation-kernel}
\end{figure}

\begin{figure}[htpb]
  \centering
  \includegraphics[width=\linewidth]{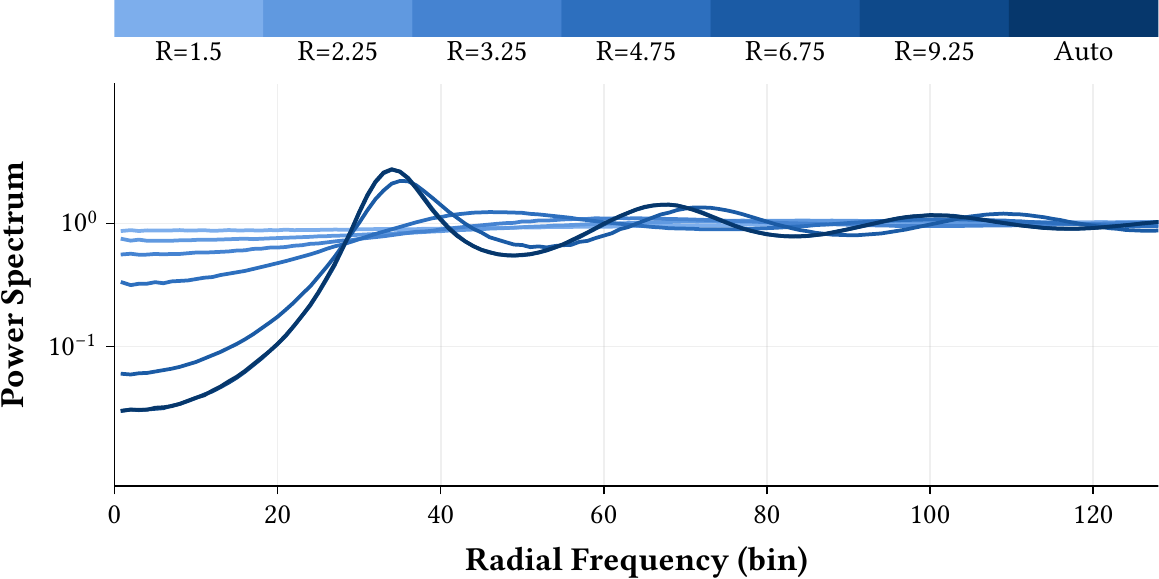}
    \caption{\textbf{Spatial-truncation ablation.} RAPS overlay for $R \in \{1.5,\allowbreak 2.25,\allowbreak 3.25,\allowbreak 4.75,\allowbreak 6.75,\allowbreak 9.25,\allowbreak 9.60\}$ with $\alpha$ recalibrated per $R$. Below $R=3.25$ no principal peak forms; $R=4.75$ is partially resolved; for $R \geq 6.75$ differences are visible only under close inspection, consistent with capturing nearly all of the kernel mass. The automatic setting ($\epsilon=10^{-6}$) yields $R=9.60$.}
  \label{fig:exp-r-ablation-raps}
\end{figure}

The kernel $k_{\sigma,p}(r) = \exp[-(r/\sigma)^p]$ has unbounded support; the truncation radius $R$ is a user-controlled parameter that determines how much interaction mass is retained, with the discarded remainder treated as a spatial tail error (Section~\ref{sec:sampler-truncations}). Figure~\ref{fig:exp-r-ablation-kernel} shows the kernel profile and tail mass as a function of $R$. Figure~\ref{fig:exp-r-ablation-raps} shows the effect on the sampled ensemble's RAPS for $R \in \{1.5, 2.25, 3.25, 4.75, 6.75, 9.25, 9.60\}$, with $\alpha$ recalibrated per $R$; the automatic setting ($\epsilon = 10^{-6}$) corresponds to $R = 9.60$, retaining over $99.99\%$ of the kernel mass.

\paragraph{Tiled execution: certified versus practical halo.}

For a tiled output to be guaranteed to agree with a full-domain run sharing the same rank field, every core pixel's dependency cone must lie entirely within the working window. Section~\ref{sec:sampler-halo} gives a certified bound: a halo of width $h \ge H_{\mathrm{det}} = D I R_{\mathrm{det}}$ guarantees this for every possible rank field.\footnote{$R_{\mathrm{det}}$ is the largest $\ell_\infty$ offset among retained edges, which bounds how far one Jacobi sweep can carry information. For a Euclidean truncation it equals the truncation radius rounded down to an integer, since the retained set contains the axis-aligned offsets up to that length.} The halo test below uses a shallow configuration ($D=2$, $I=20$, $R_{\det}=15$) for which the certified halo $H_{\mathrm{det}} = 600$ can itself be tested. This bound is conservative: errors can only enter from outside the window, and in practice dependency chains rarely reach that far.

To find a smaller working halo, we estimate the distribution of dependency-chain radii by tracing Monte Carlo chains through the sampler's read relation. Each chain starts at a core pixel, follows one randomly chosen retained edge per Gibbs step, and advances to the next depth when the visited rank exceeds the current rank or after $I$ consecutive same-layer steps; we record the maximum toroidal distance reached from the originating tile. The procedure mirrors Algorithm~\ref{alg:noisy-dependency-trace}. We then set the halo to quantiles of this distribution over $67$M chains and compare tiled against full-domain patterns at each.

\begin{figure}[t]
  \centering
  \includegraphics{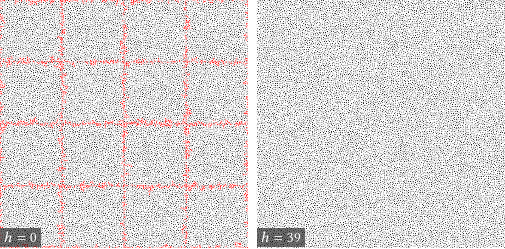}
  \caption{\textbf{Tile-boundary consistency.} Red dots mark pixels whose sampled value differs from the corresponding full-domain reference under the same rank field. Left: $h=0$, $3243$ differing pixels. Right: $h=39$, no observed differences.}
  \label{fig:exp-halo-tiled}
\end{figure}

Table~\ref{tab:exp-halo-comparison} shows that disagreement falls steadily as the halo grows and vanishes once it reaches the maximum observed chain radius, $h = 39$, far below $H_{\mathrm{det}}$. Figure~\ref{fig:exp-halo-tiled} shows where the disagreement lies: at $h = 0$, differing pixels concentrate along tile boundaries, exactly where the dependency cone first exits the window, and adding halo removes them from the outside in. Since Monte Carlo follows only a finite number of chains, each along randomly chosen edges, a rare chain that runs unusually far may never appear, so longer chains cannot be ruled out and the empirical halo is a practical rather than a certified guarantee. It is checked directly by the tiled-versus-full comparison above. The certified $H_{\mathrm{det}}$ remains the only bound that holds for every rank field, and the gap to the empirical value is the price of that generality: $H_{\mathrm{det}}$ depends only on the retained stencil, $D$, and $I$, and uses neither the kernel weights nor the realized ranks. Tighter bounds that exploit either are left to future work.

\begin{table}[t]
  \centering
  \begin{tabular*}{\columnwidth}{@{\extracolsep{\fill}} l r r @{}}
    \toprule
    Chain-radius quantile & Halo $h$ (px) & Differing pixels \\
    \midrule
    $p_{50}$ and $p_{90}$ & $0$  & $3243$ \\
    $p_{95}$           & $1$  & $2564$ \\
    $p_{99}$           & $6$  & $1311$ \\
    $p_{99.5}$         & $8$  & $875$  \\
    $p_{99.9}$         & $12$ & $429$  \\
    $p_{99.99}$        & $18$ & $90$   \\
    Max                & $39$ & $\mathbf{0}$ \\
    \midrule
    Certified ($D I R_{\mathrm{det}}$) & $600$ & $\mathbf{0}$ \\
    \bottomrule
  \end{tabular*}
  \caption{\textbf{Disagreement falls with halo size and vanishes by the maximum observed chain radius, far below the certified halo.} Halo sizes are set to quantiles of the dependency-chain radius, estimated from $67$M Monte Carlo chains at $\sigma=6.2$, $p=6$, $R=15$, $D=2$, $I=20$ (mean radius $0.21$). Differing pixels are counted between tiled and full-domain patterns on a $1024\times1024$ periodic domain at $\beta=12$, cut into $256\times256$ core tiles, both using the same rank field, with all rows from a single run using one calibrated $\alpha$. The maximum is an empirical bound: longer chains may exist outside the sampled set. The certified halo $h_\mathrm{cert} = D I R_{\mathrm{det}}$ guarantees zero disagreement for every rank field; in this configuration it is $600$.} \label{tab:exp-halo-comparison}
\end{table}

\paragraph{Empirical time complexity.}

\begin{figure}[htpb]
  \centering
  \includegraphics[width=\linewidth]{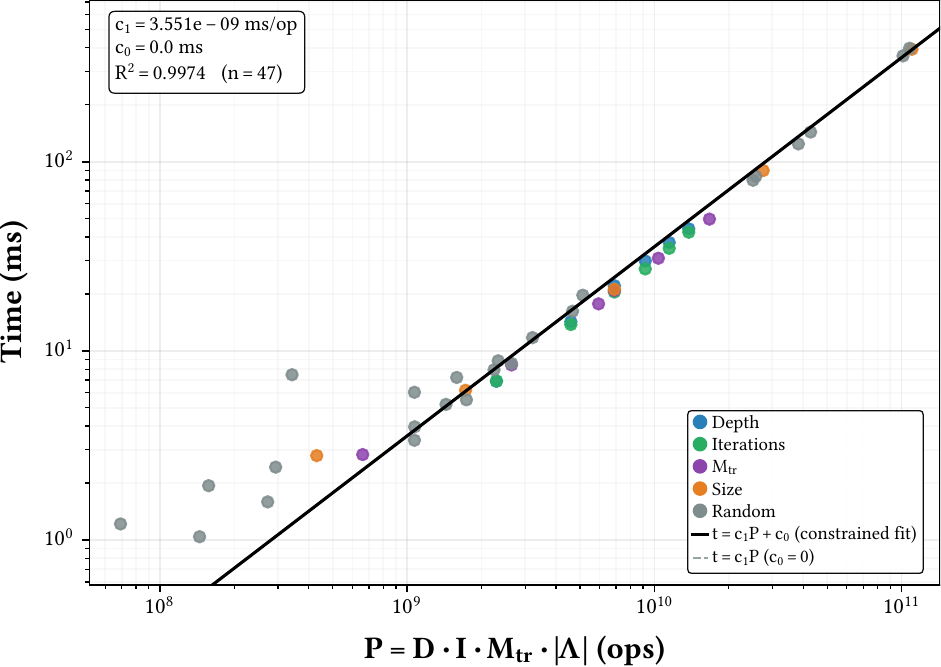}
  \caption{\textbf{Empirical time complexity.}  Wall-clock time $t$ versus the predicted product $P = D \cdot I \cdot M_{\mathrm{tr}} \cdot |\Lambda|$ across a parameter grid, fitted with $t = c_1 \cdot P + c_0$ by least squares with $c_0, c_1 \ge 0$.  The intercept absorbs fixed per-run overhead.  The fit has $R^2 \ge 0.99$, so run time is well described as proportional to $P$. }
  \label{fig:exp-scaling}
\end{figure}

Figure~\ref{fig:exp-scaling} verifies the predicted $O(D \cdot I \cdot M_{\mathrm{tr}} \cdot |\Lambda|)$ scaling using least-squares regression with an intercept term, $t = c_1 \cdot P + c_0$, where $P = D \cdot I \cdot M_{\mathrm{tr}} \cdot |\Lambda|$. Both coefficients are constrained to be nonnegative, since run time cannot be negative for a nonnegative amount of work. The early-exit test is disabled for this measurement, so the work actually performed matches the product it is regressed against. The intercept $c_0$ absorbs fixed per-run overhead (kernel launch, memory transfer) that is not part of the asymptotic model.  Fitting each factor alone, with the others held fixed, gives $R^2 \ge 0.99$ in every case. Across a parameter grid spanning several orders of magnitude in $P$, the combined fit also achieves $R^2 \ge 0.99$, so run time is well described as proportional to the product of the four factors, as the theoretical analysis predicts.

\begin{figure*}[t]
  \centering
  \includegraphics{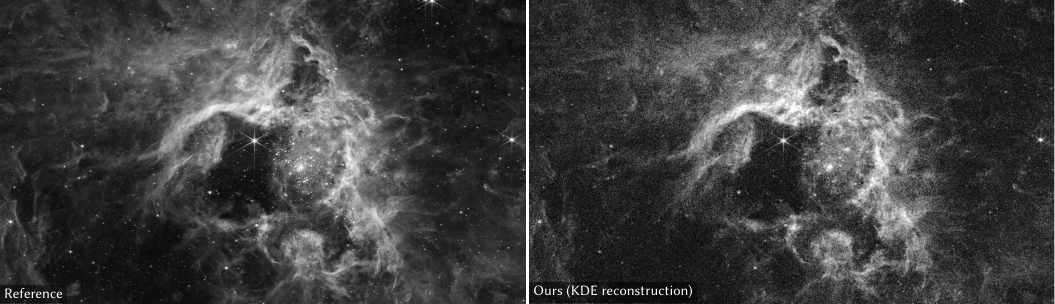}
  \caption{\textbf{Kernel density estimate (KDE) reconstruction of the adaptive stippling result.}
  Tarantula Nebula~\cite{WebbTarantula2022}.
  Left: original grayscale reference.
  Right: KDE reconstruction of the stippling result, histogram-matched to the
  source image so that remaining visual discrepancy reflects density estimation
  error rather than display normalization.
  The sampler faithfully tracks the brightness structure across more than an order
  of magnitude in local density, from the bright ionized filaments to the dark
  dust pillars.}
  \label{fig:exp-adaptive-stippling}
\end{figure*}

\begin{algorithm}[t]
\caption{Monte Carlo Dependency-Chain Tracing}
\label{alg:noisy-dependency-trace}
\begin{algorithmic}[1]
\Require Core tile $B$ on a torus of side $L$, start site $(x_0,y_0)\in B$, depth cap $D$, iteration budget $I$, weighted neighbor set $\mathcal{N}$, rank field $\{r_i\}$
\Ensure Maximum dependency radius $m$
\State $(x,y) \gets (x_0,y_0)$,\ $d \gets 0$,\ $b \gets I$
\State $m \gets d_B(x,y)$     \Comment{toroidal $\ell_\infty$ distance from $(x,y)$ to $B$, cf.\ Eq.~\eqref{eq:sampler-halo-tile}}
\While{$d < D$}
    \If{$b = 0$}
        \State $d \gets d+1$,\ $b \gets I$
    \Else
        \State Sample $(dx,dy)\in\mathcal{N}$ uniformly or by kernel weight
        \State $(x',y') \gets \left((x+dx)\bmod L,\ (y+dy)\bmod L\right)$
        \If{$r_{x',y'} < r_{x,y}$}
            \State $b \gets b-1$
        \Else
            \State $d \gets d+1$,\ $b \gets I$
        \EndIf
        \State $(x,y) \gets (x',y')$
        \State $m \gets \max\!\left(m,\ d_B(x,y)\right)$
    \EndIf
\EndWhile
\State \Return $m$
\end{algorithmic}
\end{algorithm}

\subsection{Application: Adaptive Stippling}
\label{sec:exp-stippling}

We demonstrate spatially adaptive generation on a $14557\times8418$ grayscale image of the Tarantula Nebula captured by the James Webb Space Telescope (NIRCam; NASA/ESA/CSA/STScI). The input is converted to a spatially varying target density map. Offline, we calibrate activities $\{\alpha_k\}$ for a sequence of homogeneous densities $\{\rho_k\}$ on periodic calibration patches, forming a single globally shared lookup table $A(\rho)$. For the input image, linear interpolation of this table produces the full-resolution activity map
\begin{equation}
    \alpha_i \approx A(\rho_i).
\end{equation}
During tiled generation, each tile reads the corresponding window of this global activity map; no calibration is performed per tile. Each $1024\times1024$ core tile is sampled within a $(1024+2\times128)\times(1024+2\times128)=1280\times1280$ working window, after which the halo is discarded. The adaptive configuration uses $h=128$, empirically validated by direct comparison against full-domain generation.

Figure~\ref{fig:teaser} shows the full stippling result alongside a crop of a high-density-variation region, illustrating that the sampler faithfully tracks the brightness structure of the source image across more than an order of magnitude in local density.\footnote{The dot radius is $r = \sqrt{WH/(\pi N)}$ for $N$ dots on a $W\times H$ canvas, scaled to the output resolution at render time and floored to one pixel for very small exports.} Figure~\ref{fig:exp-adaptive-stippling} shows the kernel density estimate (KDE) reconstruction of the stippling result alongside the original grayscale reference. The KDE output is histogram-matched to the source image so that any remaining visual discrepancy reflects density estimation error rather than display normalization.

Tiled execution is what keeps the memory footprint independent of the image size. Table~\ref{tab:exp-adaptive-summary} summarizes the memory footprint of each method. GBN requires $509$ MiB of GPU memory in single precision for the full $14557\times8418$ image, scaling linearly at ${\approx}4.15$ MiB/Mpx ($R^2 > 0.9999$); this is consistent with single-precision storage of a per-pixel feature map. BNOT requires substantially more RAM at large resolutions, running out of memory at $12$K resolution on our test machine. Our tiled sampler's memory footprint is determined solely by the working window size and is independent of total image resolution. The three configurations listed all use the same $128$-pixel halo, so the working window is the tile plus $256$ pixels in each dimension and memory scales with the window area rather than the tile area. Smaller tiles cost less memory but spend a larger share of the window on halo that is computed and then discarded. The mean per-tile generation time is $975$~ms; since tiles share no runtime dependencies, the total wall-clock time equals the single-tile time when sufficient compute nodes are available.

As in the homogeneous study above (Table~\ref{tab:exp-halo-comparison}, Figure~\ref{fig:exp-halo-tiled}), a validated halo makes tiled output agree with full-domain output. For the adaptive configuration used here, the $h=128$ tiled result is bitwise identical to a single-pass full-domain run over the whole $14557\times8418$ image. Since the outputs are identical, all derived statistics, including spectral observables, are identical as well.

\begin{table}[htpb]
  \centering
  \begin{tabular*}{\columnwidth}{@{\extracolsep{\fill}}llr@{}}
    \toprule
    Method & Configuration & Peak memory \\
    \midrule
    Ours & $512^2$ ($256{+}128{\times}2$) & $8.7$ MiB (VRAM) \\
    Ours & $768^2$ ($512{+}128{\times}2$) & $18.7$ MiB (VRAM) \\
    Ours\rlap{$^\star$} & $1280^2$ ($1024{+}128{\times}2$) & $50.7$ MiB (VRAM) \\
    GBN  & full image & $509$ MiB (VRAM) \\
    BNOT & full image & ${>}16$ GiB (RAM) \\
    \bottomrule
  \end{tabular*}
  \caption{\textbf{Peak memory for adaptive stippling at $14557\times8418$.}
  All Ours configurations use constant memory independent of image size;
  $^\star$denotes the configuration used for all visual results in this paper.
  GBN and BNOT memory scales with image resolution; BNOT runs out of memory at $12$K on the test machine.
  See Figure~\ref{fig:teaser} for memory scaling curves.}
  \label{tab:exp-adaptive-summary}
\end{table}

\subsection{Application: Multi-Class Blue Noise}
\label{sec:exp-multiclass}

We demonstrate a multi-class extension of the same local Gibbs construction by replacing the binary occupancy variable with a categorical state $z_i\in\{0,1,\ldots,q\}$, where $0$ denotes an empty site and $1,\ldots,q$ denote the classes. The experiment uses a shared spatial kernel and a cross-class coupling parameter $\gamma\in[0,1]$: an occupied pair of the same class receives the full repulsive penalty, whereas an occupied pair of different classes receives a fraction $\gamma$ of that penalty. The single-site update is therefore a softmax over the empty state and the $q$ class labels. At $\gamma=0$, only points of the same class repel one another; at $\gamma=1$, the repulsion depends only on total occupancy; intermediate values balance per-class and combined blue-noise structure.

The same rank-ordered DAG reconstruction applies, with the binary sigmoid update replaced by this categorical softmax. The present experiment uses the shared-kernel specialization. The supplementary material develops the Potts and multi-scale blue-noise extensions formally and proves that the structural guarantees of Section~\ref{sec:sampler} carry over: the equivalence with rank-ordered sweeps, finite-depth consistency, exactness of the Jacobi iteration at a fixed point, halo locality, and the work and storage bounds. Unlike prior constructions that derive inter-class distance constraints from geometric arguments~\cite{Wei2010MulticlassBN, Jiang2015BlueNS} or optimal-transport objectives~\cite{Salan2022}, our formulation specifies the relative same- and cross-class interactions directly through the Potts coupling.

\begin{figure}[htpb]
  \centering
  \includegraphics{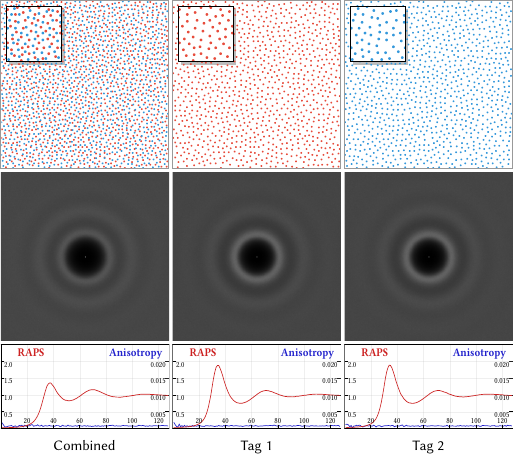}
  \caption{\textbf{Two-class blue noise.} On a $256\times256$ periodic domain.
  Columns: combined pattern, class~1, class~2.
  Rows: spatial patterns (with magnified inset), PSD, and RAPS with anisotropy.
  Both per-class and union spectra exhibit standard blue-noise structure.}
  \label{fig:exp-multiclass-two}
\end{figure}

\begin{figure}[htpb]
  \centering
  \includegraphics[width=0.75\linewidth]{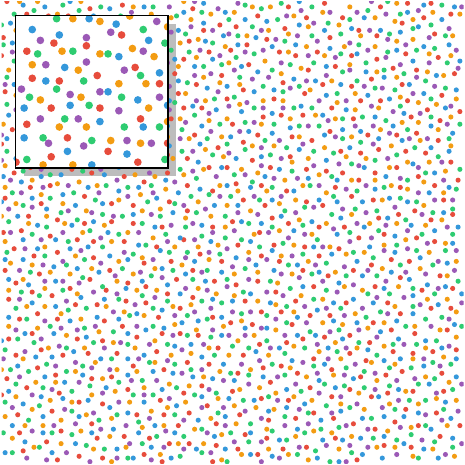}
  \caption{\textbf{Five-class blue noise.} $256\times256$ periodic domain; same DAG reconstruction as the two-class case.}
  \label{fig:exp-multiclass-five}
\end{figure}

Figure~\ref{fig:exp-multiclass-two} shows a two-class result. Each column corresponds to the combined pattern, class~1, and class~2 respectively; rows show spatial patterns, PSD, and RAPS with anisotropy. Note that on a discrete grid, adjacent pixels may belong to different classes; this is an inherent consequence of the grid spacing and does not indicate a deficiency of the construction. Figure~\ref{fig:exp-multiclass-five} applies the same Potts extension to five classes using the same DAG reconstruction, demonstrating that the construction is not restricted to the two-class case.

Combining this with the spatially varying density of Section~\ref{sec:adaptive} is a natural next step. It is supported in theory: the structural guarantees of Section~\ref{sec:sampler} carry over to the categorical update above. It requires a different calibration procedure, since the single-density search of Section~\ref{sec:sampler-adaptive} does not extend directly to several interacting classes; we leave this outside the scope of the present paper.

\section{Conclusion}
\label{sec:conclusion}

The main existing local approach to blue-noise generation, tile-based synthesis, achieves locality by replacing sampling with offline construction: point sets are prepared in advance and assembled at run time, rather than drawn from an explicit distribution. We have shown that this trade is not necessary. Modeling blue noise as an explicit Gibbs distribution with a pairwise potential on a lattice, and sampling it by truncating a backward dependency trace to a bounded spacetime cone, yields a sampler that is simultaneously local and parallel, and whose output is a draw from an explicit, finite-range and finite-depth approximation of the model. The approximation is confined to the distribution, not to the locality: with a halo at least as wide as a certified worst-case width, a tile reproduces the corresponding region of the full-domain output exactly, and the memory cost depends only on the working window rather than on the total output size. For a given configuration, the halo needed in practice can be measured directly, and in our experiments it was far smaller than this width.

\paragraph{Limitations and Future Work.}
Four design choices carry corresponding costs. First, the lattice quantizes the point positions and restricts exact spectral analysis to finite periodic domains. Second, density enters the distribution through the activity $\alpha$, so the point count is controlled only in expectation rather than fixed exactly (Section~\ref{sec:model-parameters}); this is rarely a concern in typical applications, but the count cannot be prescribed exactly. Third, the pairwise energy exposes repulsion strength, kernel scale, and kernel hardness as continuous parameters, but does not directly prescribe the resulting spectral profile; characterizing the family of profiles spanned by $(\beta,\sigma,p)$ and solving the inverse problem of matching a target spectrum remain open. Fourth, the backward truncation trades exactness for a cost and a dependency region that are known before sampling starts; in strongly-ordered regimes the footprint grows with the interaction scale and depth, and when it approaches the working domain tiling ceases to be advantageous. A natural extension is to combine the multi-class construction of Section~\ref{sec:exp-multiclass} with spatially varying density, which requires a calibration procedure for several interacting classes that does not follow directly from the single-class approach. Higher dimensions are also a natural target, though the neighborhood size, halo volume, and memory footprint all grow rapidly with dimension; representations that reduce this cost while preserving the locality guarantees of the sampler would be a worthwhile direction.

\bibliographystyle{ACM-Reference-Format}
\bibliography{reference}

\end{document}